\documentclass[final,5p,times,twocolumn]{elsarticle}
\usepackage{hyperref}
\usepackage{mathrsfs,amsmath}

\hypersetup{colorlinks=true, citecolor=blue, filecolor=blue, linkcolor=blue, urlcolor=blue}
\usepackage[utf8]{inputenc}
 
\usepackage{listings}
\usepackage{xcolor}
\usepackage{siunitx}

\definecolor{codegreen}{rgb}{0,0.6,0}
\definecolor{codegray}{rgb}{0.5,0.5,0.5}
\definecolor{codepurple}{rgb}{0.58,0,0.82}
\definecolor{backcolour}{rgb}{0.95,0.95,0.92}
 
\lstdefinestyle{mystyle}{
    backgroundcolor=\color{backcolour},   
    commentstyle=\color{codegreen},
    keywordstyle=\color{magenta},
    numberstyle=\tiny\color{codegray},
    stringstyle=\color{codepurple},
    basicstyle=\ttfamily\footnotesize,
    breakatwhitespace=false,         
    breaklines=true,                 
    captionpos=b,                    
    keepspaces=true,                 
    numbersep=5pt,                  
    showspaces=false,                
    showstringspaces=false,
    showtabs=false,                  
    tabsize=2
}
 
\usepackage{amssymb}
\usepackage{gensymb}

\newcounter{bla}

\journal{Computer Physics Communications}

\begin{document}

\begin{frontmatter}



\title{PyXtal\_FF: a Python Library for Automated Force Field Generation}


\author[a]{Howard Yanxon}
\author[a]{David Zagaceta}
\author[b]{Binh Tang}
\author[b]{David Matteson}
\author[a]{Qiang Zhu\corref{author}}

\cortext[author] {Corresponding author.\\\textit{E-mail address:} qiang.zhu@unlv.edu}
\address[a]{Department of Physics and Astronomy, University of Nevada Las Vegas, Las Vegas, NV 89154, USA}
\address[b]{Department of Statistics and Data Science, Cornell University, Ithaca, NY 14853, USA}

\begin{abstract}
We present PyXtal\_FF---a package based on Python programming language---for developing machine learning potentials (MLPs). The aim of PyXtal\_FF is to promote the application of atomistic simulations by providing several choices of structural descriptors and machine learning regressions in one platform. Based on the given choice of structural descriptors (including the atom-centered symmetry functions, embedded atom density, SO4 bispectrum, and smooth SO3 power spectrum), PyXtal\_FF can train the MLPs with either the generalized linear regression or neural networks model, by simultaneously minimizing the errors of energy/forces/stress tensors in comparison with the data from the ab-initio simulation. The trained MLP model from PyXtal\_FF is interfaced with the Atomic Simulation Environment (ASE) package, which allows different types of light-weight simulations such as geometry optimization, molecular dynamics simulation, and physical properties prediction. Finally, we will illustrate the performance of PyXtal\_FF by applying it to investigate several material systems, including the bulk SiO$_2$, high entropy alloy NbMoTaW, and elemental Pt for general purposes. Full documentation of PyXtal\_FF is available at \url{https://pyxtal-ff.readthedocs.io}.

\end{abstract}


\end{frontmatter}

\noindent
{\bf PROGRAM SUMMARY}\\
\begin{small}
\noindent
{\em Program Title:} PyXtal-FF \\
{\em Licensing provisions:} MIT \cite{1}\\
{\em Programming language:} Python 3 \\
{\em Nature of problem:} In materials modelling, the potential energy surface of a system is often computed by either the \textit{ab initio} method or an approximated classical force field. The former is proven to be the most accurate but computationally demanding, while the latter is much cheaper but suffers from insufficient accuracy. As such, a new approach to resolve the dilemma in comprising between accuracy and cost is needed. \\
{\em Solution method:} By representing the atomic structures to a set of atom-centered descriptors, one can employ a variety of machine learning models (e.g., linear regression, artificial neural networks) to effectively learn the relationship between the descriptors and energy. Due to the flexible nature of machine learning, these approaches often yield better accuracy while maintaining lower computational cost in comparison to the \textit{ab initio} method. \\
\  
\\

\end{small}

\section{Introduction}
\label{intro}
Molecular dynamics (MD) simulations have been used routinely to model the physical behaviors of many complex systems \cite{yamakov2002dislocation, terrones2002molecular, li2010dislocation}. The accuracy of the simulations is highly dependent on the underlying potential energy surface (PES) of the system. In principle, MD simulations can be based on \textit{ab initio} quantum-mechanical \cite{kresse1993ab} or classical force field methods. The \textit{ab initio} MD (AIMD) simulations usually employ density functional theory (DFT) approximation \cite{kohn1965self}, which can provide a reliable representation of the system. Despite the accuracy, DFT simulation can be extendable only to a few hundreds of atoms at a few picoseconds. This is due to solving the Kohn-Sham equation requires thousands of quantum-mechanical calculations that are scaled at $O(N^3)$ with respect to the number of atoms $N$. In consequence, simulating the structural evolution of many of complicated systems in DFT remains demanding in spite of the remarkable progress in computational facilities and efficient algorithms. This bottleneck in the DFT method is likely to persist in the foreseeable future. On the other hand, the classical MD method can model large systems at long-time scale, countering the unwavering issue of the DFT method. A great amount of efforts has been dedicated in developing PES using the classical method \cite{daw1984embedded, daw1993embedded, tersoff1986new, stillinger1985computer, mackerell1998all}. The reconstruction of the PES is usually based on simple analytical functions related to the scalar properties of the system. The class force fields can be applied to comprehend the qualitative behavior of the system. However, they are often inadequate to describe the quantitative properties of the system.

Recently, machine learning methods have been widely applied to resolve the dilemma in comprising between accuracy and cost \cite{behler2015constructing}. The machine learning potential (MLP) are trained by minimizing the cost function to attune the model to deliberately describe the \textit{ab initio} data. The cost of atomistic simulation is orders of magnitude lower than the quantum mechanical simulation, allowing the system to be scaled up to $10^5$--$10^6$ atoms \cite{artrith2012high, li2017study}.

Among many different ML models, two regression techniques are becoming increasingly popular in the materials modelling community. They include the neural networks and Gaussian process regressions. The neural networks approach has an unbiased mathematical form that can adapt to any set of reference points through an iterative fitting process given ``enough" training data. The first well accepted neural networks potential (NNP) was originally applied to elemental silicon system by Behler and Parrinello \cite{behler2007generalized}, which demonstrated that the NNP was able to reproduce the energetic sequences of many silicon phases, as well as the radial distribution function of a silicon melt at 3000 K from DFT simulation. To gain a better predictive power, they also proposed to use a series of symmetry functions (see section \ref{wACSF}), instead of the Cartesian coordinates, as the descriptors to represent the atomic environment. Since then, many attempts have been undertaken to improve the capability of neural networks approach \cite{behler2011atom}. The accomplishments of neural networks approach have been extended to multi-component \cite{artrith2011high, hajinazar2017stratified} and organic \cite{gastegger2015high} systems. In addition, Gaussian Approximation Potential (GAP), in conjunction with the bispectrum coefficients of atomic neighbour density (see section \ref{bispectrum_coefficient})), was first introduced to model the carbon, silicon, germanium, iron, and gallium nitride \cite{bartok2010gaussian}. GAP was further enhanced by replacing bispectrum coefficients with smooth overlapping power spectrum coefficients with explicit radial basis \cite{bartok2013representing}. Similar to GAP, Thompson \textit{et al.} \cite{thompson2015spectral} developed (quadratic) Spectral Neighbor Analysis Potential (SNAP) method based on the Taylor expansion of bispectrum coefficients. In addition, linear regression model based on the moment tensor---comparable to atomic environments inertia tensors---as the descriptor \cite{shapeev2016moment} was also demonstrated to be a competitive approach. Many applications based on different MLP models have shown that machine learning potentials work remarkably well in different types of atomistic simulations \cite{chen2017accurate, li2018quantum, szlachta2014accuracy, deringer2018data1, deringer2018data2, podryabinkin2019accelerating}. 

In the recent years, several software packages \cite{artrith2011high, singraber2019parallel, lee2019simple, khorshidi2016amp, PINN-2020, schutt2018schnet} were developed to train the MLPs. Among these, the RuNNer \cite{artrith2011high} is a closed source software for developing NNP, and ænet \cite{artrith2016implementation} is mainly written in FORTRAN/C and utilizes atom-centered symmetry functions (see section \ref{wACSF}) as the descriptor. Similar codes, such as the n2p2 package \cite{singraber2019parallel} in C++, SIMPLE-NN package \cite{lee2019simple} in Python/C, and AMP package \cite{khorshidi2016amp} in Python/FORTRAN, have similar feature as ænet package. SIMPLE-NN leverages the capability of Tensorflow platform---a deep learning GPU-accelerated library, and AMP provides several other descriptors such as Zernike and bispectrum components. Our recent works also suggested that NNP can be developed using bispectrum and power spectrum components as the descriptor while training on energy, forces, and stress simultaneously \cite{yanxon2020transferability, Zagaceta2020SpectralNN}. Moreover, DeepPot-SE \cite{zhang2018end} and SchNetPack \cite{schutt2018schnet} packages introduce additional filters to the descriptor such as distance-chemical-species-dependent filter and continuous convolutional filter, respectively, prior to the deep learning model.

In this paper, we present PyXtal\_FF---an open-source package in Python scripting language---for developing MLP such as NNP and generalized linear potential (GLP). The objective of PyXtal\_FF is to provide handy user-interface in developing MLP with training of energy, force, and stress contributions simultaneously. PyXtal\_FF creates MLP based on atom-centered descriptors such as (weighted) atom-centered symmetry functions \cite{behler2015constructing}, embedded atom density \cite{zhang2019embedded}, SO(4) bispectrum coefficients \cite{bartok2010gaussian}, and smooth SO(3) power spectrum \cite{bartok2013representing}. Finally, we will demonstrate the usage of the current features of the package with $\textrm{SiO}_2$ \cite{lee2019simple}, high entropy alloy \cite{li2019unravelling}, and elemental Pt \cite{zhang2018end} as examples.


\section{Theory}
\label{theory}

In this section, we will provide in-depth discussions of the two main ingredients in creating MLP: atom-centered descriptor and regression technique. The construction of the total energy of a crystal structure can be written as the collections of atomic energy contributions, in which is a functional ($\mathscr{E}$) of the atom-centered descriptor ($\boldsymbol{X}_i$):
\begin{equation}\label{total_E}
    E_\textrm{total} = \sum_{i=1}^{N} E_i = \sum_{i=1}^{N} \mathscr{E}_i(\boldsymbol{X}_i)
\end{equation}
Specifically, the functional represents regression techniques such as neural networks or generalized linear regressions.

Since neural networks and generalized linear regressions have well-defined functional forms, the analytic derivatives can be derived by applying the chain rule to obtain the force at each atomic coordinate, $\boldsymbol{r}_m$:
\begin{equation}\label{force}
     \boldsymbol{F}_m=-\sum_{i=1}^{N}\frac{\partial \mathscr{E}_i(\boldsymbol{X}_{i})}{\partial \boldsymbol{X}_{i}} \cdot \frac{\partial
    \boldsymbol{X}_{i}}{\partial \boldsymbol{r}_m}
\end{equation}
Force is an important property to accurately describe the local atomic environment especially in geometry optimization and MD simulation. Finally, the stress tensor is acquired through the virial stress relation:
\begin{equation}\label{stress}
     \boldsymbol{S}=-\sum_{m=1}^N \boldsymbol{r}_m \otimes \sum_{i=1}^{N} \frac{\partial \mathscr{E}_i(\boldsymbol{X}_{i})}{\partial \boldsymbol{X}_{i}} \cdot \frac{\partial
    \boldsymbol{X}_{i}}{\partial \boldsymbol{r}_m},
\end{equation}
where $\otimes$ is the outer product.

According to Eqs. \ref{force} and \ref{stress}, one needs to compute the energy derivative $\frac{\partial \mathscr{E}} {\partial \boldsymbol{X}}$ and the derivatives of descriptor $X$ with respect to the atomic positions. For a structure with $N$ atoms and $L$ descriptors per atom, the energy derivative is a 2D array of [$N, L$]. The force related derivative (dxdr) can be best organized as a 4D array with the dimension of [$N, N, L, 3$]. Note that dxdr[$i, j, :, :$] is zero when the $i$-$j$ atomic pair has a distance larger than the cutoff distance. Thus, it may become a sparse array when the structure has a large number of atoms. Correspondingly, one can easily derive the 5D rdxdr array by multiplying $r$ to each dxdr according to the outer product. In Python, one can simply compute the forces and stresses based on the following Einstein summation.

\begin{lstlisting}[language=Python, 
caption=Force and stress computation in Python.,
label={c0}]
import numpy as np

"""
Einstein summation to compute force and stress.
dedx: 2D array [N, L]
dxdr: 4D array [N, N, L, 3]
rdxdr: 5D array [N, N, L, 3, 3]
force: 2D array [N, 3]
stress: 2D array [3, 3]
"""
force = -np.einsum("ik, ijkl->jl", dedx, dxdr)
stress = -np.einsum("ik, ijklm->lm", dedx, rdxdr)

\end{lstlisting}

\subsection{Atom-centered Descriptors}\label{atom-center-descriptors}

Descriptor---a representation of a crystal structure---plays a critical role in constructing reliable MLP. If the MLP is directly mapped from the atomic positions or the Cartesian coordinates, it can only describe systems with the same number of atoms due to the fixed length of the regression input. In addition, Cartesian coordinates are poor descriptors in describing the structural environment of the system, restricted by the periodic boundary conditions. While the total energy of the structure remains the same by translation, rotational, or permutation operations, the atomic positions will change. Several types of descriptors have been developed in the past few years \cite{HIMANEN2020106949}. For example, Coulomb matrix has been widely used due to its simplicity. Coulomb matrix encompasses self interaction based on the nuclear charge and Coulomb repulsion between two nuclei \cite{rupp2012fast, faber2015crystal}. Logically, the Coulomb matrix can be upgraded for periodic crystals through Ewald summation that includes long range interaction calculated in reciprocal space. In addition, many-body tensor representation---derives from Coulomb matrix while related to bag of bonds which corresponds to different types of bonding in molecular systems---can be used for both finite and periodic systems when interpretability/visualization is desirable \cite{huo2017unified}. These descriptors have been widely used to model the molecules.

In the atom-centered descriptors, one usually needs to consider the neighboring environment for the centered atom within a cutoff radius of $R_c$. To ensure the descriptor mapping from the atomic positions smoothly approaching zero at the $R_c$, a cutoff function ($f_c$) is included to every mapping scheme:
    \begin{equation} \label{cosine}
    f_c(R) = \begin{cases}
        \frac{1}{2}\cos\left(\pi \frac{R}{R_c}\right) + \frac{1}{2} & R \leq R_c\\
        0              & R > R_c
    \end{cases}
    \end{equation}
where $R$ is distance. The cutoff function is zero at $R_c$ and the intensity decreases as $R$ approaches $R_c$. Consequently, the derivative of the cutoff function is:
    \begin{equation} \label{cosineprime}
    \frac{\partial f_c}{\partial R} = \begin{cases}
        -\frac{\pi}{2R_c}\sin\left(\pi \frac{R}{R_c}\right) & R \leq R_c\\
        0              & R > R_c
    \end{cases}
    \end{equation}
One should heed of the importance of the vanishing derivative of cutoff function at $R_c$, which is important in describing the force. By definition, there is no discontinuity as the slope decays to zero at $R_c$.

In the following, we will introduce four types of atom-centered descriptors in details. The corresponding derivative terms can be found in \ref{dACSF}, \ref{dEAD} and our recent work \cite{Zagaceta2020SpectralNN}.

\subsubsection{(Weighted) Atom-centered Symmetry Functions ($G$)}\label{wACSF}

The atom-centered symmetry functions (ACSFs) are the very first types of descriptors used in the MLP development \cite{behler2007generalized}. In general, there are two classes of ACSFs: radial and angular symmetry functions \cite{behler2015constructing}. The radial symmetry function or $G^{(2)}$ describes the radial distribution of the atomic environment, and the angular symmetry functions, $G^{(4)}$ and $G^{(5)}$, account for the three-body angular distribution of atoms in the neighborhood. The $G^{(2)}$ is expressed as the sum of the radial distances between the center atom $i$ and the neighbor atoms $j$ as follow:
\begin{equation}\label{G2}
G^{(2)}_i = \sum_{j\neq i} e^{-\eta (R_{ij}-R_s)^2} \cdot f_c(R_{ij})
\end{equation}
Here, $G^{(2)}$ value is controlled by the width ($\eta$) and the shift ($R_s$).

$G^{(4)}$ and $G^{(5)}$ symmetry functions are a few of many ways to capture the angular information via three-body interactions ($\theta_{ijk}$). As the structures are constraint by the periodic boundary condition, a three-body periodic description such as $\cos(\theta_{ijk})$ is used. The explicit form of $G^{(4)}$ and $G^{(5)}$ are:

\begin{equation} \label{G4}
    \begin{split}
    G^{(4)}_i = &2^{1-\zeta}\sum_{j\neq i} \sum_{k \neq i, j} [(1+\lambda \cos \theta_{ijk})^{\zeta} \cdot e^{-\eta (R_{ij}^2 + R_{ik}^2 + R_{jk}^2)} \cdot \\
        &f_c(R_{ij}) \cdot f_c(R_{ik}) \cdot f_c(R_{jk})]
    \end{split}
\end{equation}
\begin{equation} \label{G5}
    \begin{split}
    G^{(5)}_i = &2^{1-\zeta}\sum_{j\neq i} \sum_{k \neq i, j} [(1+\lambda \cos \theta_{ijk})^{\zeta} \cdot e^{-\eta (R_{ij}^2 + R_{ik}^2)} \cdot \\
    &f_c(R_{ij}) \cdot f_c(R_{ik})]
    \end{split}
\end{equation}
$\zeta$ determines the strength of angular information. The degree of $\zeta$ is normalized by $2^{1-\zeta}$ for unvarying the values of $G^{(4)}$ and $G^{(5)}$ symmetry functions due to ranges of $\zeta$. $\lambda$ values are set to +1 and -1, for inverting the shape of the cosine function. The difference between $G^{(4)}$ and $G^{(5)}$ symmetry functions is in the interactions between the neighbors $j$ and $k$. The modification in $G^{(5)}$ symmetry function yields in dampening value of $G^{(5)}$, which can be beneficial in representing larger atomic separation between the two neighbors.

Clearly, the number of ACSFs will grow depending on chemical species as the separations of chemical species are needed. For instance, in a binary AB system, the number of $G^{(2)}$ ACSFs on specie A need to double to distinguish A-A and A-B pair interactions. For $G^{(4)}$, three different triplets A-A-A, A-A-B, B-A-B (where the middle position denotes the center atom) will be needed. To avoid this unpleasant growth, one can apply a weighting parameter based on the chemical species when counting these atomic pairs and triplets. One popular choice is simply to use the atomic number as the weighting parameters. Hence, Gastegger and coauthors proposed the weighted version of ACSF \cite{gastegger2018wacsf}, in which each component of the radial and angular symmetry functions in Eqs. (\ref{G2}, \ref{G4}, \ref{G5}) can be multiplied by the followings:
\begin{gather*}
\textrm{the weighted ACSF:~~~~~} 
\begin{cases}
Z_j & \textrm{radial}\\
Z_j Z_k & \textrm{angular}
\end{cases}
\end{gather*}
where $Z_j, Z_k$ represents the atomic number of neighboring atom $j$ and $k$.

To obtain a satisfactory MLP model, one has to choose a set of parameters to construct the (w)ACSF descriptors, which may require some demanding human intervention \cite{gastegger2018wacsf, zhang2018end, schutt2018schnet}. As mentioned, the choice of $\lambda$ is straightforward. In general, $\zeta$ takes the value of 1. Increasing $\zeta$ focuses on the strength of the angular information in region close to \ang{0} and \ang{180}, and decreasing it will weaken the contribution of angular information at around \ang{90}. Since the exponential term has larger effect on the symmetry functions, the selection of $\eta$ and $R_s$ can be more elaborate. Several routines are available in the literature \cite{gastegger2018wacsf, zhang2019embedded} by fixing $\eta$ while varying $R_s$ or vice versa.

\subsubsection{Embedded Atom Density ($\rho$)}

Embedded atom density (EAD) descriptor \cite{zhang2019embedded} is inspired by embedded atom method (EAM)---description of atomic bonding by assuming each atom is embedded in the uniform electron cloud of the neighboring atoms \cite{daw1984embedded, daw1983semiempirical}. In EAD, the electron density is modified by including the square of the linear combination the atomic orbital components:
\begin{equation}\label{ead}
    \rho_i(R_{ij}) = \sum_{l_x, l_y, l_z}^{l_x+l_y+l_z=L_{\max}} \frac{L_{\max}!}{l_x!l_y!l_z!} \bigg(\sum_{j\neq i}^{N} Z_j  \Phi(R_{ij})\bigg)^2
\end{equation}
where $Z_j$ represents the atomic number of neighbor atom $j$. $L_{max}$ is the quantized angular momentum, and $l_{x,y,z}$ are the quantized directional-dependent angular momentum. For example, $L_{max}=2$ corresponds to the $d$ orbital. Lastly, the explicit form of $\Phi$ is:
\begin{equation}\label{PHI}
    \Phi(R_{ij}) = \frac{x^{l_x}_{ij}  y^{l_y}_{ij}  z^{l_z}_{ij}}{R_c^{l_x+l_y+l_z}} \cdot e^{-\eta (R_{ij}-R_s)^2} \cdot f_c(R_{ij})
\end{equation}
According to quantum mechanics, $\rho$ follows the similar procedure in determining the probability density of the states, i.e. the Born rule.

EAD can be regarded as an alternative version of ACSF without classification between the radial and angular term. The angular or three-body term is implicitly incorporated in when $L_{\max}>0$ \cite{zhang2019embedded}. By definition, the computation cost for calculating EAD is cheaper than angular symmetry functions by avoiding the extra sum of the $k$ neighbors. In term of usage, the parameters $\eta$ and $R_s$ are similar to the strategy used in the Gaussian symmetry functions, and the maximum value for $L_{max}$ is 3, i.e. up to $f$ orbital.

\subsubsection{SO(4) Bispectrum ($B$)}\label{bispectrum_coefficient}

The SO(4) bispectrum components \cite{bartok2010gaussian,bartok2013representing,thompson2015spectral} are another type of atom-centered descriptor based on the harmonic analysis of the atomic neighbor density function on the 3-sphere.
The atomic neighbor density function is given by \cite{bartok2013representing}:
\begin{equation}\label{density}
    \rho(\boldsymbol{r}) = \delta(\boldsymbol{r}) + \sum_i^{R_{\textrm{c}}} w_i  f_c(\boldsymbol{r}_{i})  \delta(\boldsymbol{r}-\boldsymbol{r_i}) 
\end{equation}
Where $w_i$ is a species dependent weight factor and $f_c$ is a cutoff function. The cutoff function is $f_c$ is introduced to ensure that the atomic neighbor density function goes smoothly to zero at the cutoff.  

Then we map the atomic neighbor density function from 3-D euclidean space to another 3-D space, the surface of a four dimensional hypersphere:
\begin{equation*}
    \begin{split}
        s_1 &= r_0\cos\omega \\
        s_2 &= r_0\sin\omega\cos\theta \\
        s_3 &= r_0\sin\omega\sin\theta\cos\phi \\
        s_4 &= r_0\sin\omega\sin\theta\sin\phi,
    \end{split}
\end{equation*}
where $r_0$ is a parameter and the polar angles are defined by:
\begin{equation}\label{polar}
    \begin{split}
        \theta &= \arccos\left(\frac{z}{r}\right)\\
        \phi &= \arctan\left(\frac{y}{x}\right)\\
        \omega &= \frac{\pi r}{r_0}
    \end{split}
\end{equation}

The Winger-D matrix elements ($D^j_{m',m}$) are the harmonic functions on the 3-sphere, therefore an arbitrary function defined on the 3-sphere can be expanded in terms of Wigner-D matrix elements.  Here we expand the atomic neighbor density function on the 3-sphere in terms of Wigner-D matrices.
\begin{equation*}
    \rho(\boldsymbol{r}) = \sum_{j=0}^{+\infty}\sum_{m',m = -j}^{+j}{c^j_{m',m}D^j_{m',m}\left(\omega;\theta,\phi\right)}
\end{equation*}

Where the expansion coefficients $c^j_{m',m}$ are given by the following inner product

\begin{equation}
    c^j_{m',m} = \left<D^j_{m',m}|\rho(\boldsymbol{r})\right> =  D^{*j}_{m',m}(\boldsymbol{0}) + \sum_i^{r_i \leq R_{\textrm{c}}} f_{\textrm{c}}(r_i)D^{*j}_{m',m}(\omega_i;\theta_i,\phi_i)
\end{equation}
Finally, the SO(4) bispectrum components can then be calculated using third order products of the expansion coefficients:
\begin{equation}
    \begin{split}
    B_{j_1,j_2,j} &= \sum_{m',m = -j}^{j}c^{*j}_{m',m}\sum_{m_1',m_1 = -j_1}^{j_1}c^{j_1}_{m_1',m_1}\times \\ & \sum_{m_2',m_2 = -j_2}^{j_2}c^{j_2}_{m_2',m_2}C^{jj_1j_2}_{mm_1m_2}C^{jj_1j_2}_{m'm_1'm_2'},
    \end{split}
\end{equation}
where $C$ is a Clebsch-Gordan coefficient.

\subsubsection{Smooth SO(3) Power Spectrum ($P$)}
The Smooth SO(3) Power Spectrum components were been proposed to describe the atomic local environment \cite{bartok2013representing}. In contrast to the SO(4) bispectrum components, the Smooth SO(3) power spectrum is based on an alternative atomic neighbor density while also expanded on the 2-sphere and a radial basis.  The alternative atomic neighbor density is defined in terms of Gaussians as follows:
\begin{equation}\label{Gaussdensity}
    \rho '(\boldsymbol{r}) = \sum_i^{r_i \leq R_{\textrm{c}}} w_i e^{-\alpha|\boldsymbol{r}-\boldsymbol{r_i}|^2},
\end{equation}
Then the atomic neighbor density function is then expanded in terms of spherical harmonics and a radial basis $g_n(r)$
as shown in Eq. \ref{Gaussdensity}:
 
 \begin{equation*}
     \rho'(\boldsymbol{r}) = \sum_{l=0}^{+\infty}\sum_{m=-'l}^{+l}c_{nlm}g_n(r)Y_{lm}(\boldsymbol{\hat{r}})
 \end{equation*}
 
 Where the expansion coefficients $c_{nlm}$ are given by
 
\begin{equation*}
\begin{split}
     c_{nlm} &= \left<Y_{lm}g_n(r)|\rho'\right> = \\ &  4\pi \sum_i^{r_i \leq R_{\textrm{c}}} w_i e^{-\alpha r_i^2}  Y^*_{lm}(\boldsymbol{\hat{r}_i})\times \\ &\int_0^{R_{c}}r^2g_n(r)I_l(2\alpha r r_i)e^{-\alpha r^2}dr
\end{split}
\end{equation*}
where $I_l$ is a modified spherical Bessel function of the first kind.  A convenient radial basis for this purpose, $g_n(r)$, consisting of cubic and higher order polynomials, orthonormalized on the interval $(0, R_{c})$ has been suggested by Bartok \cite{bartok2013representing}.  

\begin{equation}
    g_n(r) = \sum_\alpha W_{n,\alpha}\phi_\alpha(r)
\end{equation}

where $W_{n,\alpha}$ are the orthonormalization coefficients given by the relation to the overlap matrix $\boldsymbol{S}$ by  $\boldsymbol{W} = \boldsymbol{S}^{-1/2}$ and

\begin{equation*}
    \phi_\alpha(r) = (R_{c}-r)^{\alpha+2}/N_\alpha
\end{equation*}

\begin{equation*}
    N_\alpha = \sqrt{\frac{2r_{\textrm{cut}}^{(2\alpha+7)}}{(2\alpha+5)(2\alpha+6)(2\alpha+7)}}
\end{equation*}

And the elements of the overlap matrix $\boldsymbol{S}$ are given by

\begin{equation}
    \begin{split}
       & S_{\alpha\beta} = \int_0^{r_{\textrm{cut}}}r^2\phi_\alpha(r)\phi_\beta(r)dr \\ &= 
        \frac{\sqrt{(2\alpha+5)(2\alpha+6)(2\alpha+7)(2\beta+5)(2\beta+6)(2\beta+7)}}{(5+\alpha+\beta)(6+\alpha+\beta)(7+\alpha+\beta)}
    \end{split}
\end{equation}

and finally, the Smooth SO(3) power spectrum is given by

\begin{equation}
    p_{n_1 n_2 l} = \sum_{m=-l}^{+l}c_{n_1lm}c^*_{n_2lm}
\end{equation}

\subsection{Regression Models}
Here, we discuss the regression model, i.e., the functional form ($\mathscr{E}$) presented in Eq. \ref{total_E}. Each regression model is species-dependent, i.e. as the the number of species increases, the regression parameters will increase. For the sake of simplicity, we will explanation the regression models for the  single-species system.

In any regression model, the objective is to minimize a loss function which describes the discrepencies between the prediction and true reference values (including energy, force, and stress tensors) for each atomic configuration in the training data set. 
\begin{equation}\label{loss}
\begin{split}
    \Delta = \frac{1}{2M}\sum_{i=1}^M\Bigg[\bigg(&\frac{E_i - E^{\textrm{Ref}}_i}{N_{\textrm{atom}}^i}\bigg)^2 +
             \frac{\beta_f} {3N_{\textrm{atom}}^i}\sum_{j=1}^{3N_{\textrm{atom}}^i}
    (F_{i, j} - F_{i, j}^{\textrm{Ref}})^2 \\
    &+ \frac{\beta_s} {6} \sum_{p=0}^{2} \sum_{q=0}^{p}
    (S_{pq} - S_{pq}^{\textrm{Ref}})^2 \Bigg]
\end{split}
\end{equation}
where $M$ is the total number of structures in the training pool, and $N^{\textrm{atom}}_i$ is the total number of atoms in the $i$-th structure. The superscript $\textrm{Ref}$ corresponds to the target property. $\beta_f$ and $\beta_s$ are the force and stress coefficients respectively. They scale the importance between energy, force, and stress contribution as the force and stress information can overwhelm the energy information due to their sizes. Additionally, a regularization term can be added to induce penalty on the entire parameters preventing overfitting:
\begin{equation}\label{penalty}
    \Delta_\textrm{p} = \frac{\alpha}{2M} \sum_{i=1}^{m} (\boldsymbol{w}^i)^2
\end{equation}
where $\alpha$ is a dimensionless number that controls the degree of regularization.

Clearly, one has to choose differentiable functional as well as its derivative due to the existence of force ($F$) and stress ($S$) contribution along with the energy ($E$) in the loss function. In the following sections, generalized linear regression and neural network regression will be introduced.

\subsubsection{Generalized Linear Regression}

This regression methodology is a type of polynomial regression. Essentially, the quantum-mechanical energy, forces, and stress can be expanded via Taylor series with atom-centered descriptors as the independent variables:
\begin{equation}\label{PolyEnergy}
    E_{\textrm{total}} = \gamma_0 + \boldsymbol{\gamma} \cdot \sum^{N}_{i=1}\boldsymbol{X}_i + \frac{1}{2}\sum^{N}_{i=1}\boldsymbol{X}_i^T\cdot \boldsymbol{\Gamma} \cdot \boldsymbol{X}_i + \cdots
\end{equation}
where $N$ is the total atoms in a structure. $\gamma_0$ and $\boldsymbol{\gamma}$ are the weights presented in scalar and vector forms. $\boldsymbol{\Gamma}$ is the symmetric weight matrix (i.e. $\boldsymbol{\Gamma}_{12} = \boldsymbol{\Gamma}_{21}$) describing the quadratic terms. In this equation, we only restricted the expansion up to polynomial 2 due to to enormous increase in the weight parameters.

In consequence, the force on atom $j$ and the stress matrix can be derived according to Eqs. (\ref{force}, \ref{stress}), respectively:
\begin{equation}\label{PolyForce}
    \boldsymbol{F}_m = -\sum^{N}_{i=1} \bigg(\boldsymbol{\gamma} \cdot \frac{\partial \boldsymbol{X}_i}{\partial \boldsymbol{r}_m} + 
    \frac{1}{2} \bigg[\frac{\partial \boldsymbol{X}_i^T}{\partial \boldsymbol{r}_m} \cdot \boldsymbol{\Gamma} \cdot \boldsymbol{X}_i + \boldsymbol{X}_i^T \cdot \boldsymbol{\Gamma} \cdot \frac{\partial \boldsymbol{X}_i}{\partial \boldsymbol{r}_m} \bigg]\bigg)
\end{equation}

\begin{equation}\label{PolyStress}
    \boldsymbol{S} = -\sum_{m=1}^N \boldsymbol{r}_m \otimes \sum^{N}_{i=1} \bigg(\boldsymbol{\gamma} \cdot \frac{\partial \boldsymbol{X}_i}{\partial \boldsymbol{r}_m} + 
    \frac{1}{2} \bigg[\frac{\partial \boldsymbol{X}_i^T}{\partial \boldsymbol{r}_m} \cdot \boldsymbol{\Gamma} \cdot \boldsymbol{X}_i + \boldsymbol{X}_i^T \cdot \boldsymbol{\Gamma} \cdot \frac{\partial \boldsymbol{X}_i}{\partial \boldsymbol{r}_m} \bigg]\bigg)
\end{equation}

Note that the energy, force, and stress share the weights parameters $\{\gamma_0, \boldsymbol{\gamma}_1, ..., \boldsymbol{\gamma}_N, \boldsymbol{\Gamma}_{11}, \boldsymbol{\Gamma}_{12}, ..., \boldsymbol{\Gamma}_{NN}\}$. 
Once the energy, force and stress tensors are known, the derivative of the loss function can be evaluated. Finding the zero derivative of loss function (Eq. \ref{loss}) in linear regression is equivalent to solve a set of linear equations of $\boldsymbol{Ax} = \boldsymbol{b}$. In PyXtal\_FF, we construct such $\boldsymbol{A}$ matrix and use the
\textit{numpy.linalg.lstsq} solver to obtain the least-squares solution. 

\subsubsection{Neural Network Regression}
\begin{figure}[ht]
\centering
\includegraphics[width=0.40\textwidth]{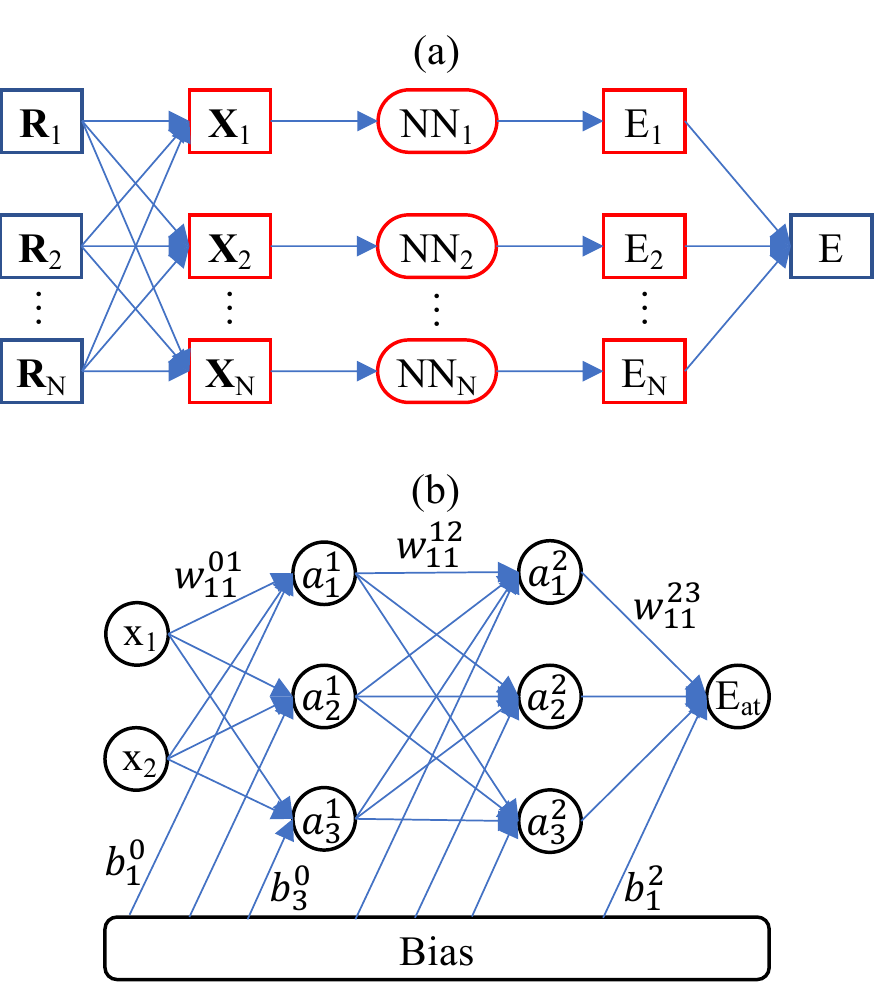}
\caption{(a) A schematic diagram of high-dimensional neural networks. (b) A zoom-in version of the color-coded part in (a).}
\label{HDNN}
\end{figure}

Compared to the linear regression, neural networks provides more flexible functionals to fit a large data sets. Figure \ref{HDNN} shows a schematic diagram based on neural networks training. Prior to the neural networks architecture, the atom-centered descriptors are mapped based on the atomic environment of a structural configuration as discussed in the previous section. These descriptors serve as the input to the neural networks architecture and are arranged in the first layer as shown in Figure \ref{HDNN}b. The next layers are the hidden layers. Neural networks can simply cast more weights parameters as needed through increasing number of hidden layers and/or hidden layers nodes without the increasing number of descriptors. The nodes in the hidden layers carry no physical meaning. 

For each of the hidden nodes, activation functions such as Tanh and Sigmoid functions are frequently used in our NNP implementation. While ReLU as an activation function is extremely popular in image processing, we believe ReLU is not an appropriate choice in constructing MLP, due to the function carries discontinuity at zero. These nodes are connected via the weights and biases and propagate in forward direction only. In the end, the output node represents the atomic energy. A mathematical form to determine any node value can be written as:
\begin{equation}\label{neuron}
    X^{l}_{n_i} = a^{l}_{n_i}\bigg( b^{l-1}_{n_i} + \sum^{N}_{n_j=1} W^{l-1, l}_{n_j, n_i} \cdot X^{l-1}_{n_j} \bigg)
\end{equation}
The value of a neuron ($X_{n_i}^l$) at layer $l$ can determined by the relationships between the weights ($W^{l-1, l}_{n_j, n_i}$), the bias ($b^{l-1}_{n_i}$), and all neurons from the previous layer ($X^{l-1}_{n_j}$). $W^{l-1, l}_{n_j, n_i}$ specifies the connectivity of neuron $n_j$ at layer $l-1$ to the neuron $n_i$ at layer $l$. $b^{l-1}_{n_i}$ represents the bias of the previous layer that belongs to the neuron $n_i$. These connectivity are summed based on the total number of neurons ($N$) at layer $l-1$. Finally, an activation function ($a_{n_i}^l$) is applied to the summation to induce non-linearity to the neuron ($X_{n_i}^l$).$X_{n_i}$ at the output layer is equivalent to an atomic energy, and it represents an atom-centered descriptor at the input layer. The collection of atomic energy contributions are summed to obtain the total energy of the structure. At the end, the total energy, forces ans stress tensors are compared to the reference values (see Eq. \ref{loss}). This process is called forward propagation.

Similar to the linear regression, one needs to obtain a set of weight parameters to minimize the loss function. In NN architecture, the gradient of loss with respect to the weight parameters can be conveniently done by the backpropagation algorithm. Hence, a number of optimization algorithms can be applied here to update the weights iteratively, until the optimal solution is found.

\section{PyXtal\_FF Workflow}
\label{code}

\begin{figure}[ht]
\centering
\includegraphics[width=0.45\textwidth]{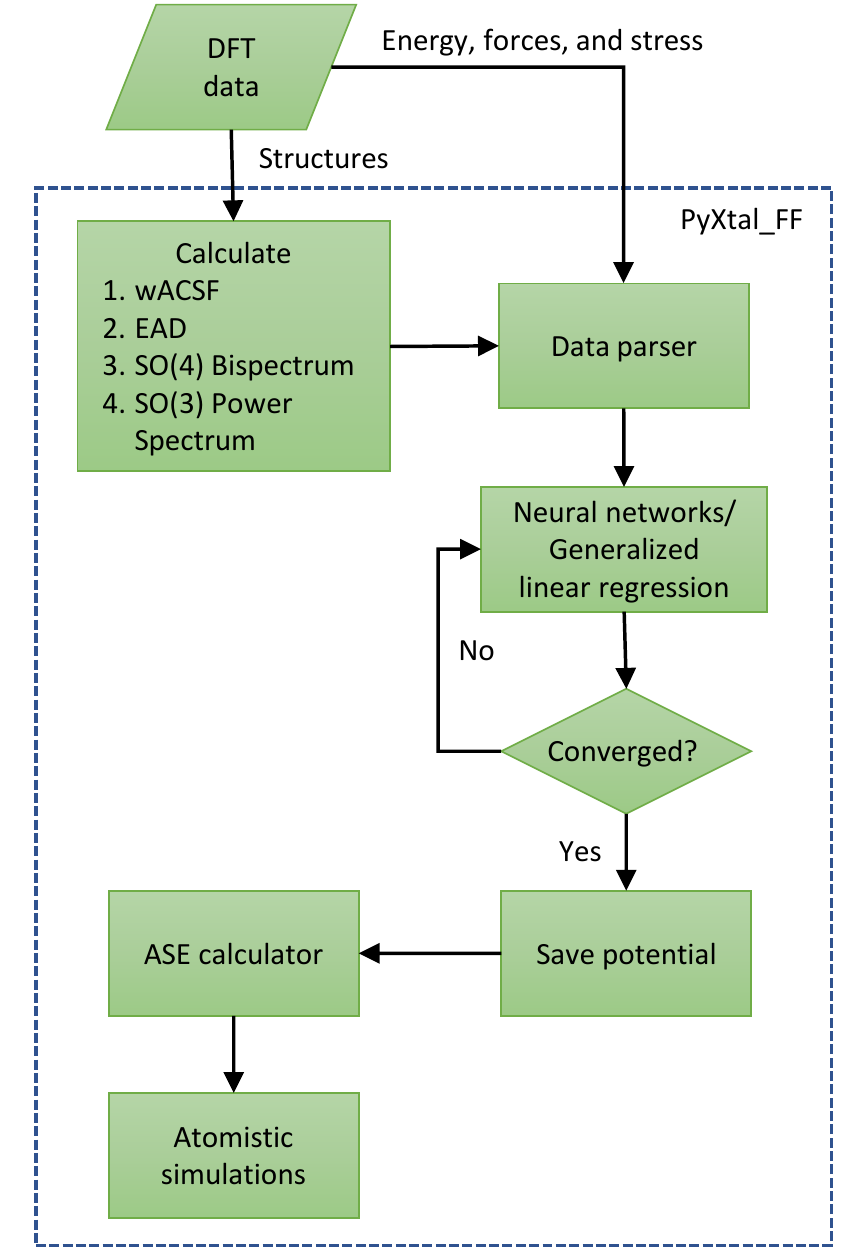}
\caption{Schematic diagram of PyXtal\_FF workflow for NNP/GLP training. In the neural network, data parser includes normalization of the calculated descriptors.}
\label{workflow}
\end{figure}

In this section, we discuss about development of PyXtal\_FF and its philosophy. PyXtal\_FF is written in Python. Presently, the package is equipped with two regression models and four types atom-centered descriptor, as explained in Section \ref{theory}. These regression models and atom-centered descriptors are easily extendable without changing the core user-interface features. Figure \ref{workflow} represents the workflow of PyXtal\_FF.

First, PyXtal\_FF utilizes the Atomic Simulation Environment (ASE) package \cite{larsen2017atomic} to parse the DFT data assembled in several formats, including ASE database, JSON, extended XYZ and the VASP OUTCAR formats. Further, ASE is also employed to compile the atomic neighborhood of each atom in the unit cell based on the periodic boundary conditions within a cutoff radius. After the neighboring data are gathered, it will compute the user-defined type of descriptor. The computation follows the theory described in Section \ref{atom-center-descriptors} utilizing NumPy---a Python library for scientific computing \cite{walt2011numpy}. For every structure, the descriptor calculator will return the descriptors and the force and stress related derivatives. Eventually, the descriptors represent as the independent variables in the regression models to obtain the energy, and the derivative terms are needed to compute the force and stress values.

For the regression, the Pyxtal\_FF supports two models, linear (quadratic) regression and neural networks. Here we focus on the latter since it is a more popular workforce for MLP development. The neural network regression is powered by PyTorch \cite{NEURIPS2019_9015_torch}---an open-source deep learning framework based on automatic differentiation \cite{paszke2017automatic}. Currently, we support three optimization algorithms for training: Limited Broyden-Fletcher-Goldfarb-Shanno (L-BFGS) \cite{liu1989limited}, adaptive Moment Estimation (Adam) \cite{kingma2014adam}, and stochastic gradient descent (SGD) with momentum \cite{qian1999momentum}. The L-BFGS method with approximated line search is the recommended optimizer when the training data is relatively small, as the quasi-Newton method is generally more stable and finds local optima more efficiently. With larger training datasets, however, the L-BFGS method is memory demanding, and one can seek to use first-order methods such as Adam or SGD with momentum. Both SGD and Adam algorithms are usually done in mini-batches, where the the gradients for each weight update are calculated based on a subset of the entire training data set. Training in mini batches can reduce the variance of the parameter updates leading to stable convergence. If needed, the training can also be done in graphical processing units (GPU) mode. 

In addition to the force field generation, PyXtal\_FF also provides the supports to utilize the trained models for several types of atomistic simulations, including geometry optimization, MD simulation, physical properties prediction, and phonon calculation. These features are managed by ASE calculator, in which the MLP potential passes the energy, forces, and stress tensors to the calculator and ASE performs the relevant atomistic simulations. Since these simulation will be powered by Python, we only recommend to use them for light weight simulations. In the near future, we are going to work on interfacing the trained MLP with LAMMPS \cite{LAMMPS} to enable the truly large scale atomistic simulation. 

\section{Example Usage}
\label{usage}
PyXtal\_FF can be used as stand-alone library in Python scripts. A PyXtal\_FF example code to train Pt model is shown in the following listing

\begin{lstlisting}[language=Python, 
caption=PyXtal\_FF script for force field training,
label={c1}]
from pyxtal_ff import PyXtal_FF

# define the path of train/test data
train = 'train.json'
test = 'test.json'

# define the descriptor
descriptor = {'type': 'Bispectrum',
              'parameters': {'lmax': 3},
              'Rc': 4.9}
              
# define the regression model              
model = {'system' : ['Pt'],
         'hiddenlayers': [16, 16],
         'epoch': 1000,
         'path': 'Pt-Bispectrum/'
         'optimizer': {'method': 'lbfgs'}}

# define the pyxtal_FF model and train it         
mlp = PyXtal_FF(descriptor, model)
mlp.run(TrainData=train, TestData=test)


\end{lstlisting}

The atom-centered descriptors and the model are described in dictionary. The dictionary keys determine the necessary command for the code and are made as intuitive as possible. Most of keys follow the hyperparameters in the section \ref{theory}. By default, PyXtal\_FF will use neural networks as the regression algorithm. Here, PyXtal\_FF will look for \textit{train.json} and \textit{test.json} files as the training and test data set, respectively. 

After the training is complete, the trained model is saved in the result folder (\textit{Pt-Bispectrum}) with a name of \textit{16-16-checkpoint.pth}, in which 16-16 denotes two hidden layers with 16 nodes each. PyXtal\_FF provides a built-in interface with the ASE code \cite{larsen2017atomic}, in which one can use the model to perform different types of calculations through ASE. Below is a simple example to perform the geometry optimization on a Pt bulk crystal (\textit{Pt\_bulk.cif}) based on the trained model from the listing \ref{c1}.

\begin{lstlisting}[language=Python, caption=PyXtal\_FF script to perform geometry optimization], label={c2}]
from pyxtal_ff import PyXtal_FF
from pyxtal_ff.calculator import PyXtalFFCalculator, optimize
from ase.io import read

# load the trained model
mliap = "Pt-Bispectrum/16-16-Pt_cluster.pth"
ff = PyXtal_FF(model={'system': ["Pt"]}, 
               logo=False)
ff.run(mode='predict', mliap=mliap)
calc = PyXtalFFCalculator(calc)

# read the structure
pt_bulk = read('Pt_bulk.cif')

# perform the relaxation 
pt_bulk.set_calculator(calc, box=True)
pt_bulk = optimize(pt_bulk)
print('energy: ', pt_bulk.get_potential_energy())

\end{lstlisting}

In addition to geometry optimization and MD simulation, PyXtal\_FF also provides several utility functions to simulate the elastic and phonon properties, which are based on several external Python libraries including Phonopy \cite{phonopy}, seekpath \cite{HINUMA2017140} and matscipy \cite{matscipy}. More detailed examples can be found in the online documentation \url{https://pyxtal-ff.readthedocs.io}.


\section{Applications}\label{examples}
In this section, we choose three different examples to illustrate the power of PyXtal\_FF and benchmark the performances of different descriptors. While the linear regression scheme is also supporte by PyXtal\_FF, we will focus on the NNP model as it provides more flexibility. The examples to be investigated mainly differ by the source of datasets, including (1) single SiO$_2$ from pure MD simulation; (2) collective data set of NbMoTaW from various approaches; (3) elemental Pt consisting of bulk, surfaces and clusters from different runs of MD simulations.

\subsection{Binary System}

The $\textrm{SiO}_2$ data set \cite{lee2019simple} was generated by the DFT method within the framework of VASP \cite{Vasp-PRB-1996}, using the generalized gradient approximation Perdew-Burke-Ernzerhof (PBE) exchange-correlation functional \cite{PBE-PRL-1996}. The kinetic energy cutoff was set to 500 eV, and the energy convergence criterion is within 10 meV/atom. The MD trajectories are taken at different temperatures including liquid, amorphous and crystalline ($\alpha$-quartz, $\alpha$-cristobalite, and tridymite) configurations. The original data set contain 3,048 $\textrm{SiO}_2$ configurations (60 atoms per structure). For simplicity, we considered a subset that consists of 1,316 structures. with the goal of to gaining an overview of performances and computation costs for each descriptor. Below gives the parameters to define each descriptor.

\begin{lstlisting}[language=Python, caption=PyXtal\_FF script to define the descriptors.], label={cd}]
# ACSF (70)
para = {'G2': 
        {'eta': [0.003214, 0.035711, 
                 0.071421, 0.124987, 
                 0.214264, 0.357106, 
                 0.714213, 1.428426],
         'Rs': [0]},
        'G4': 
        {'lambda': [-1, 1], 
          'zeta':[1, 2, 4], 
          'eta': [0.000357, 0.028569, 0.089277]}
        }
descriptor = {'type': 'ACSF',
              'Rc': 4.9, 
              'parameters': para,
             }
             
# wACSF (26)
descriptor = {'type': 'wACSF',
              'Rc': 4.9, 
              'parameters': para,
             }
             
# EAD (30)
para = {'eta': [0.003214, 0.035711, 
                0.071421, 0.124987, 0.214264],
        'Rs': [0, 1.50],
        'lmax': 2,
       }

descriptor = {'type': 'EAD',
              'Rc': 4.9, 
              'parameters': para,
             }

# SO3 (40)
descriptor = {'type': 'SO3',
              'Rc': 4.9, 
              'parameters': {'nmax': 4, 
                             'lmax': 3},
             }

# SO4 (30)
descriptor = {'type': 'SO4',
              'Rc': 4.9, 
              'parameters': {'lmax': 3},
             }
\end{lstlisting}

In short, we choose a universal cutoff value of 4.9 \AA~ for all descriptors. Each descriptors requires some manual selection of hyperparameters in the real (e.g., $\eta, \lambda, \zeta, R_s$) or integer ($l_{\max}$, $n_{\max}$) space. The ACSF parameters were taken from Ref. \cite{lee2019simple} which lead to 70 descriptors. In its wACSF version, the number is reduced to 26. For EAD, we chose a similar set of parameters for $\eta$ and $R_s$, which make 30 descriptors when $L_{\max}$ = 2. For SO3 and SO4, only the integer type hypeparameters need to be provided. In this work, we set 40 SO3 descriptors with $n_{\max}$ = 4 and $l_{\max}$ =3, and 30 SO4 descriptors with $l_{\max}$ =4. The neural network regression will be used with two hidden layers with 30 nodes each.

Table \ref{tSiO2} summarizes the performances of each training after 12000 steps. First, the ACSF-70 set yields the best accuracy in both energy (1.3 meV/atom) and forces (81.2 meV/\AA), while the errors in its corresponding wACSF-26 set rise by 60-70\% in both energy (2.1 meV/atom) and forces (141.8 meV/\AA). On the other hand, the weighted EAD-30 descriptor, supposed to mimic G2 and G4 ACSFs, gives the highest errors (4.0 meV/atom for energy and 300 meV/\AA~ for forces). This may be due to lack of optimization on the hyperparameters. However, it should be noted that the computation of EAD is much faster than ACSF. Therefore, it is worth exploring a systematic approach to obtain the optimum set for EAD. For the two spectral descriptors, SO3-40 seems to outperform SO4-30 while it cost about a similar level of CPU time. In terms of accuracy, SO3-40 (1.4 meV/atom in energy MAE and 115.1 meV/\AA~ in force MAE) is in the middle of ACSF-70 and wACSF-26. Another remarkable advantage of the spectral descriptors is that tuning the hyperparameters is much easier. If one does not want to spend too much time on choosing the hyperparameters, SO3 seems to be a better choice than ACSF. We note that all descriptor computations are based on Python. It is expected that the speed will be much faster when they are implemented in FORTRAN or C languages.  
\begin{table}[ht] 
  \centering
  \caption{The MAE values of the predicted energy and forces of 1316 SiO$_2$ data set from the 30-30 neural network models with different descriptors within 12000 L-BFGS steps of training. For each type of descriptors, the average CPU time for descriptor computation per structure is also given.}
  \begin{tabular}{lcccc}
    \hline\hline
             & CPU time   & Energy     & Force     \\
             & (secs/60 atoms)  & (meV/atom) & (meV/\AA) \\
    \hline
    ACSF (70)     & 4.374        & 1.3 & 81.2\\
    wACSF (26)    & 4.372        & 2.1 & 141.8\\
    EAD (30)     & 0.584        & 4.8 & 259.0\\
    SO3 (40)      & 1.028        & 1.4 & 115.1\\
    SO4 (30)      & 1.078        & 3.3 & 204.2\\
    \hline\hline
  \end{tabular}\label{tSiO2}
\end{table}

\subsection{High Entropy Alloy}

High entropy alloys (HEAs) are systems that encompass four or more equimolar/near-equimolar alloying elements. It has been shown that HEAs carry many interesting properties such as high hardness and corrosion resistance \cite{yeh2004nanostructured, senkov2011mechanical}. Due to the high computational cost of DFT method, HEA serves as a great example in MLP development with PyXtal\_FF. Here, we will use NbMoTaW HEA as an example \cite{li2019unravelling}, which are comprised of elemental, binary, ternary, and quaternary  systems. Each of the elemental systems has their ground state, strain-distorted, surface, and AIMD configurations. The binary alloys are composed of solid solution structures with the size of 2 $\times$ 2 $\times$ 2 supercell. Lastly, 300 K, 1000 K, and 3000 K AIMD configurations along with special quasi-random structures establish the ternary and quaternary data points. 
The total structures used in developing the MLP are 5529 configurations for training set and 376 configurations for test set.

\begin{figure}[ht]
\centering
\includegraphics[width=0.45\textwidth]{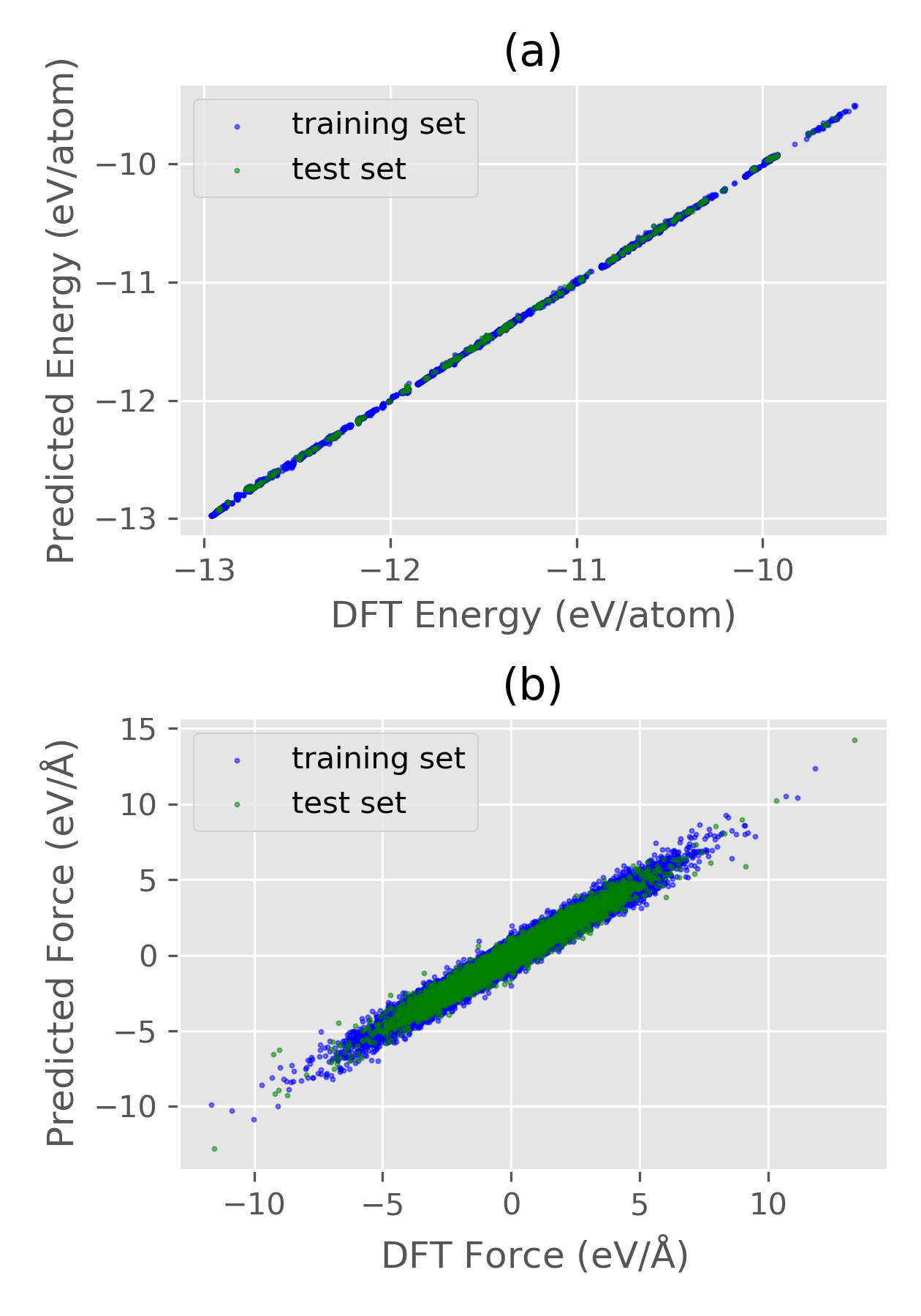}
\caption{The correlation plots between NNP and DFT for (a) energy and (b) forces in the HEA system. The energy MAE values are 6.69 meV/atom and 7.57 meV/atom for training and test sets, while the force MAE values are 0.14 eV/{\AA} and 0.17 eV/{\AA}.}
\label{NNP-HEA}
\end{figure}

In the original work \cite{li2019unravelling}, SNAP based on the linear regression predicted that the MAE values for energies are 4.3 meV/atom and 5.1 meV/atom for the training and test sets, and the MAE values in forces are 0.13 eV/{\AA} and 0.14 eV/{\AA}. The results demonstrated a quite satisfactory accuracy comparable to the quantum calculation. However, it needs to be noted that the energy training in Ref. \cite{li2019unravelling} was based on the comparison of formation energy relative to the elemental solids, which spans from -0.193 to 0.934 eV/atom for the entire dataset, whereas the atomic energy spans from -12.960 to -9.502 eV/atom. Training with the energy in a normalized range can surely reduce the error of fitting. However, this fitting method does not fully solve the force field prediction problem since it relies on some DFT reference data. We attempted to employ the linear regression model to fit only the absolute DFT energy based on the same descriptor as used in Ref. \cite{li2019unravelling}, the resulting MAE values are 944 meV/atom and 6.329 eV/$\AA$ for energy and force when the force coefficient is $10^{-4}$. The MAE values of formation energy fitting yield remarkable improvement of 22 meV/atom and 0.243 eV/$\AA$ for energy and force, respectively. Despite this improvement, the accuracy is insufficient. Perhaps, it is due to lack of fine tuning of hyperparameters, such as atomic weights and cutoff radii for each species. 

To obtain a better accuracy, we decided to fit the absolute DFT energy and forces based on the NNP model. We employed the smooth SO(3) power spectrum as the descriptor, which are formed by $n_{\max}$=4 and $l_{\max}$=3 with 40 components in total up to the cutoff radius of 5.0 \AA. The NNP training is executed with 2 hidden layers with 20 nodes for each layer while energy, force, and stress contributions are trained simultaneously. The importance coefficients of force and stress are set to $10^{-3}$ and $10^{-4}$, respectively. The results of the NNP training is illustrated in Figure \ref{NNP-HEA}. The NNP energy MAE values for the training and test sets are 6.69 meV/atom and 7.57 meV/atom, and the NNP force MAE values are 0.14 eV/{\AA} and 0.17 eV/{\AA}. In addition, the MAE value of stress for training set is 0.078 GPa. Our results of energy and force yield worse performance compared to the previous report. Nevertheless, our NNP model offers a more general representation of the DFT PES since it does not rely on any prior reference values.

Furthermore, we calculated physical properties such as elastic constants, bulk and shear moduli, and the Poisson's ratio of the cubic elemental crystals (see Table \ref{elastic_table}). From the table, the overall performances of NNP in predicting the physical properties are reasonable, except that the $C_{44}$ value of Nb is negative. However, this is consistent with the fact that the DFT's $C_{44}$ is also significantly lower than other terms. Hence, the negative $C_{44}$ is acceptable if one considers the noise of stress data in training. Meanwhile, the $C_{44}$ value may be remedied by providing additional training data set focusing on the shearing effect of Nb, increasing the importance of the stress coefficient, or increasing the hidden layer size in the NNP training. In addition, SO(3) descriptor can be conveniently expanded in terms of both radial basis ($n_{\max}$) and angular momentum ($l_{\max}$) for achieving better overall accuracy.

\begin{table}[ht]
\caption{Comparison of physical properties predicted with SO(3)-NNP. The DFT and experiment values are obtained from Ref \cite{li2019unravelling}. $B$ and $G$ denote the empirical Voigt-Reuss-Hill average bulk and shear moduli. $\nu$ is the Poisson's ratio. The DFT results were taken from open database of Materials Project \cite{MP-2013}.}
\centering
\begin{tabular}{lccccccc}
\hline
\hline
 &$C_{11}$ & $C_{12}$ & $C_{44}$ & $B$ & $G$  & $\nu$ \\ \hline
 & (GPa) & (GPa) & (GPa) & (GPa) & (GPa) &  \\ \hline
 
 Mo & & & & & & & \\
 DFT & 472 & 158 & 106 & 262 & 127 & 0.30 \\
 Ref\cite{li2019unravelling} & 435 & 169 & 96 & 258 & 110 & 0.31 \\ 
 NNP & \textbf{453} & \textbf{161} & \textbf{107} & \textbf{259} & \textbf{121} & \textbf{0.30} \\ \hline
 
 Nb & & & & & & & \\
 DFT & 233 & 145 & 11 & 174 & 24 & 0.45 \\
 Ref\cite{li2019unravelling} & 266 & \textbf{142} & \textbf{20} & 183 & \textbf{32} & 0.42 \\
 NNP & \textbf{255} & 130 & -3 & \textbf{171} & N/A & \textbf{0.47} \\ \hline
 
 Ta & & & & & & & \\
 DFT & 265 & 158 & 69 & 194 & 63 & 0.35 \\
 Ref\cite{li2019unravelling} & \textbf{257} & \textbf{161} & \textbf{67} & \textbf{193} & 59 & 0.36 \\
 NNP & 280 & 165 & 72 & 203 & \textbf{66} & \textbf{0.35} \\ \hline
 
 W & & & & & & & \\
 DFT & 510 & 201 & 143 & 304 & 147 & 0.29 \\
 Ref\cite{li2019unravelling} & 560 & 218 & 154 & 332 & 160 & \textbf{0.29} \\
 NNP & \textbf{527} & \textbf{196} & \textbf{143} & \textbf{306} & \textbf{151} & \textbf{0.29} \\\hline\hline

\end{tabular}\label{elastic_table}
\end{table}

\subsection{Pt MLP for General Purposes}

Compared to crystalline systems, surfaces and nanoparticles generally represent the more challenging cases in MLP training as the nanoparticle contain more versatile atomic environments and more complex PES is expected. Here, we applied the NNP model to a Pt data set \cite{zhang2018end}, which consists of three data types: Pt surface, Pt bulk, and Pt cluster. There are 927 clusters of 15 atoms, and the Pt bulk type consists of 1717 configurations which are composed of 256 atoms. Pt surface are constructed from (001), (110), and (111) surfaces. Respectively, there are 949, 819, and 700 structures which consist of 320, 160, and 320 atoms. In our NNP development, we chose 90\% of the total structures randomly as the training set, and the remaining 10\% as the test set. The SO(3) power pectrum descriptor with $l_{\max}=3$ and $n_{\max}=4$ at radius cutoff of 4.9 $\AA$ was used to construct the MLP in the NNP model with two hidden layers with 30 nodes each. Unlike the previous examples, the minibatch scheme with the Adam optimizer was employed. In each iteration, the training process was updated in a batch size of 25 configurations. 

\begin{table}[ht]
  \centering
  \caption{The trained RMSE values of the predicted energy and forces of Pt data set from the 30-40-40-1 NNP model. For reference, the results from the DeepPot-SE model \cite{zhang2018end} is also reported. It should be noted that that DeepPot-SE results were based on training the entire MoS$_2$/Pt dataset.}
  \begin{tabular}{lcccc}
    \hline\hline
            &\multicolumn{2}{c}{SO(3)-NNP} & \multicolumn{2}{c}{DeepPot-SE \cite{zhang2018end}} \\
            &  Energy & Force & Energy  & Force \\
            & (meV) & (meV/\AA) & (meV) & (meV/\AA)\\

    \hline
    Bulk  &1.64 & 64 & 2.00 & 84 \\
    Surface & 5.91  & 87 & 6.77 & 105 \\
    Cluster & 7.63  & 247 & 30.6 & 201 \\
    \hline\hline
  \end{tabular}
  \label{nnp_pt_cluster}
\end{table}

\begin{figure}[ht]
\centering
\includegraphics[width=0.45\textwidth]{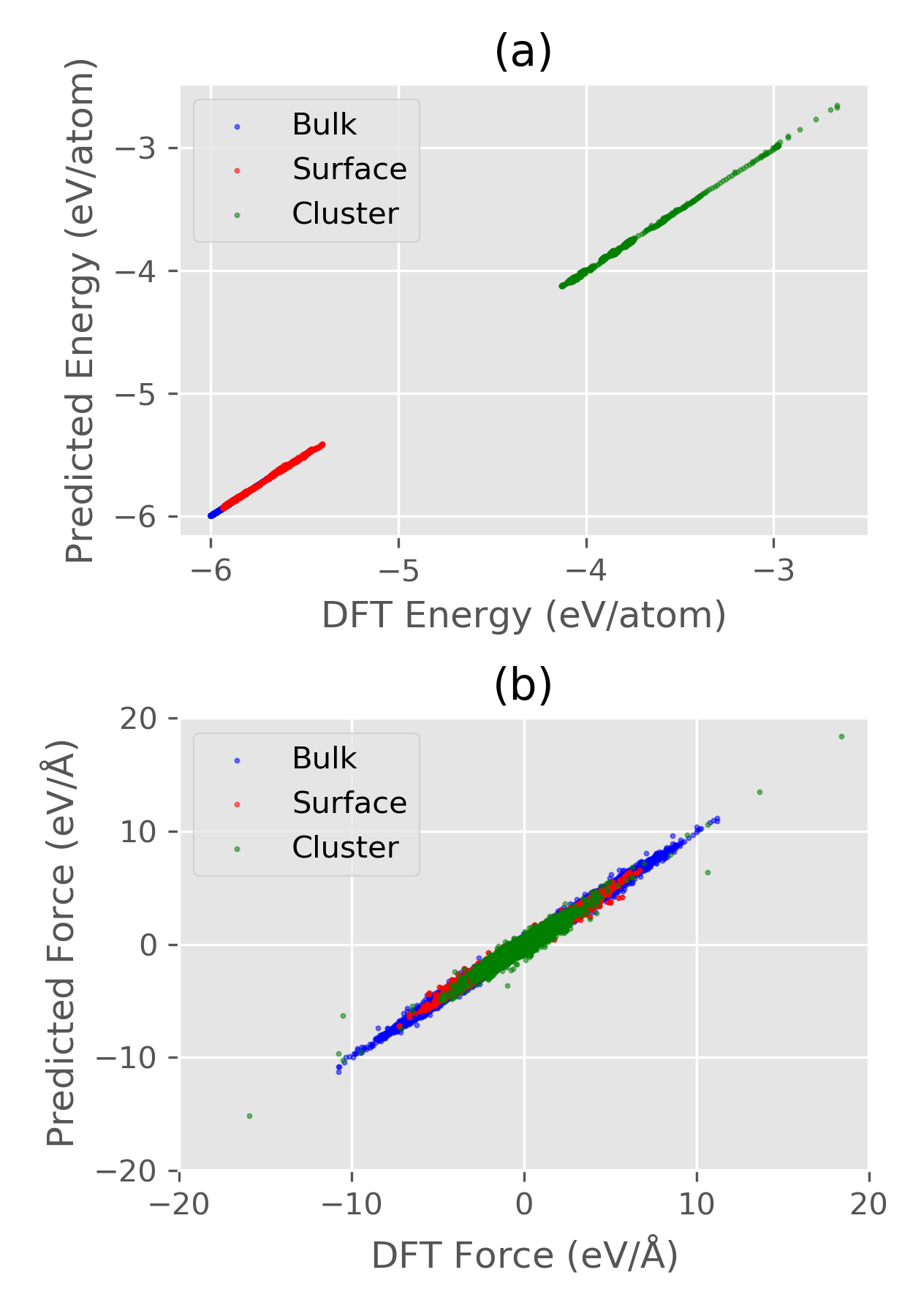}
\caption{The correlation plots between NNP and DFT for (a) energy and (b) forces in the Pt system.}
\label{Pt-HEA}
\end{figure}

In the original literature \cite{zhang2018end}, DeepPot-SE includes $\textrm{MoS}_2$ slab and Pt clusters on $\textrm{MoS}_2$ substrate ($\textrm{MoS}_2$/Pt). The performance of DeepPot-SE yields satisfactory results. Meanwhile, embedded atom neural networks method can achieves outstanding results using a fraction of the same data set \cite{zhang2019embedded}. Both of the methods exploit a large number of neural networks parameters, in the order of $10^4$--$10^5$, while the current study only adopts 2191 weight parameters for exemplary purpose. As shown in Table \ref{nnp_pt_cluster}, the accuracy from our small neural network model is comparable to that of DeepPot-SE results. Not surprisingly, the group of Pt bulk has the lowest errors with only 1.64 meV/atom for the RMSE in energy and 67 meV/\AA~ in force. On the contrary, the errors on Pt clusters are about 2-4 time higher for both energy and forces. This is expected since the local atomic environments in the clusters are more diverse and thus learning the relation is harder. Nevertheless, the values from this exploratory study is comparable to the results from deep learning models. This example also suggests that a small NNP model with the properly constructed features can be a complementary solution for MLP development.

\section{Conclusion}
\label{conclusion}

In conclusion, we introduced PyXtal\_FF a versatile package for developing MLPs that can perform at the DFT level. Currently, the code allows one to construct the MLPs from four different types of atom-centered descriptors: (w)ACSFs, EAD, SO4 bispectrum, or SO3 power spectrum. Two regression models, the generalized linear regression and neurual networks, are supported to train the MLP by simultaneously fitting the data of energy, forces and stresses from the ab-initio simulation. In particular, we focus on the neural networks potential development. Our software package utilizes PyTorch as the main machinery, which is equipped with neural network models, automatic differentiation, as well as various optimization algorithms. We demonstrated the features of the current PyXtal\_FF version by three examples on SiO$_2$, NbMoTaW HEA, and elemental Pt, respectively. In general, the mean absolute error values of each trained MLPs fall into the range of several meVs/atom in energy and several hundred 
meV/{\AA} in forces. While training on stress is optional, it is helpful to improve the general accuracy of the model. More importantly, this is crucial to yield better prediction on materials' elastic properties. As such, the MLPs can be applied to investigate the materials properties in greater accuracy than the classical potentials built from the empirical model. Finally, the PyXtal\_FF is an open source code. We welcome anyone who is interested in MLP development to contribute to this project.

\section*{Acknowledgments}
We acknowledge the NSF (I-DIRSE-IL: 1940272) and NASA (80NSSC19M0152) for their financial supports. The computing resources are provided by XSEDE (TG-DMR180040).

\appendix



\section{The Derivatives of ACSF}\label{dACSF}

Following the Eq. \ref{G2} the derivative with respect to an atom $m$ can be written in the following form:
\begin{equation}\label{G2prime}
\frac{\partial G^{(2)}_i}{\partial \boldsymbol{r}_{m}} = \sum_{j\neq i} e^{-\eta (R_{ij}-R_s)^2}  \bigg(\frac{\partial f_c}{\partial R_{ij}} - 2\eta (R_{ij}-R_s)f_c\bigg) \frac{\partial R_{ij}}{\partial \boldsymbol{r}_{m}}
\end{equation}

For the periodic system, the computation of $\frac{\partial R_{ij}}{\partial \boldsymbol{r}_{m}}$ is straightforward except that one needs to consider one additional case. When $i=j$, the derivative is always zero.
\begin{equation}\label{dRijdRm_norm}
    \frac{\partial R_{ij}}{\partial \boldsymbol{r}_{m}} =
    \begin{cases}
        0 & m \notin [i, j] \\
        0   & m = i = j \\
        -\frac{\boldsymbol{r}_{ij}}{Rij} & m = i ~(\textrm{when~} i \neq j) \\ 
        \frac{\boldsymbol{r}_{ij}}{Rij} &  m = j ~(\textrm{when~} i \neq j)\\
    \end{cases}
\end{equation}

In Eqs. (\ref{G4}, \ref{G5}), the cosine function can be defined as:
\begin{equation}\label{cos}
    \cos \theta_{ijk} = \frac{\boldsymbol{r}_{ij} \cdot \boldsymbol{r}_{ik}}{R_{ij}R_{ik}}
\end{equation}
where $\boldsymbol{r}_{ij}$ is the relative position between atom $j$ and atom $i$.

In the following, the expressions for the derivative with respect to an interacting atom $m$ are:
\begin{equation}
\begin{aligned}
&\frac{\partial G^{(4)}_i}{\partial \boldsymbol{r}_{m}} =  2^{1-\zeta} \sum_{j\neq i} \sum_{k \neq i, j} e^{-\eta (R_{ij}^2 + R_{ik}^2 + R_{jk}^2)} \\
& \Bigg[\lambda \zeta (1+\lambda \cos \theta_{ijk})^{\zeta-1} \frac{\partial \cos \theta_{ijk}}{\partial \boldsymbol{r}_{m}} f_c(R_{ij}) f_c(R_{ik}) f_c(R_{jk}) \\
& -2 \eta (1+\lambda \cos \theta_{ijk})^{\zeta} \bigg(R_{ij} \frac{\partial R_{ij}}{\partial \boldsymbol{r}_{m}}+ R_{ik} \frac{\partial R_{ik}}{\partial \boldsymbol{r}_{m}}+R_{jk} \frac{\partial R_{jk}}{\partial \boldsymbol{r}_{m}}\bigg) f_c(R_{ij}) f_c(R_{ik}) f_c(R_{jk}) \\
& + (1+\lambda \cos \theta_{ijk})^{\zeta} \bigg(\frac{\partial f_c(R_{ij})}{\partial R_{ij}} \frac{\partial R_{ij}}{\partial \boldsymbol{r}_{m}} f_c(R_{ik}) f_c(R_{jk}) \\
& + f_c(R_{ij}) \frac{\partial f_c(R_{ik})}{\partial R_{ik}} \frac{\partial R_{ik}}{\partial \boldsymbol{r}_{m}} f_c(R_{jk}) + f_c(R_{ij}) f_c(R_{ik}) \frac{\partial f_c(R_{jk})}{\partial R_{jk}} \frac{\partial R_{jk}}{\partial \boldsymbol{r}_{m}}\bigg)\Bigg]
\end{aligned}
\end{equation}

\begin{equation}
\begin{aligned}
& \frac{\partial G^{(5)}_i}{\partial \boldsymbol{r}_{m}} =  2^{1-\zeta} \sum_{j\neq i} \sum_{k \neq i, j} e^{-\eta (R_{ij}^2 + R_{ik}^2)} \\
& \Bigg[\lambda \zeta (1+\lambda \cos \theta_{ijk})^{\zeta-1} \frac{\partial \cos \theta_{ijk}}{\partial \boldsymbol{r}_{m}} f_c(R_{ij}) f_c(R_{ik}) \\
& -2 \eta (1+\lambda \cos \theta_{ijk})^{\zeta} \bigg(R_{ij} \frac{\partial R_{ij}}{\partial \boldsymbol{r}_{m}}+ R_{ik} \frac{\partial R_{ik}}{\partial \boldsymbol{r}_{m}}+R_{jk} \frac{\partial R_{jk}}{\partial \boldsymbol{r}_{m}}\bigg) f_c(R_{ij}) f_c(R_{ik}) \\
& + (1+\lambda \cos \theta_{ijk})^{\zeta} \bigg(\frac{\partial f_c(R_{ij})}{\partial R_{ij}} \frac{\partial R_{ij}}{\partial \boldsymbol{r}_{m}} f_c(R_{ik}) + f_c(R_{ij}) \frac{\partial f_c(R_{ik})}{\partial R_{ik}} \frac{\partial R_{ik}}{\partial \boldsymbol{r}_{m}} \bigg)\Bigg]
\end{aligned}
\end{equation}

The derivatives of atomic distances, $R_{ij}$ and $R_{ik}$, carry the same meaning as in Eq. \ref{dRijdRm_norm}. The expression of the cosine of triple-atom angle is
\begin{equation}
    \frac{\partial \cos \theta_{ijk}} {\partial \boldsymbol{r}_m} = \frac{\boldsymbol{r}_{ik}}{R_{ij} R_{ik}} \cdot \frac{\partial \boldsymbol{r}_{ij}}{\partial \boldsymbol{r}_m}  + \frac{\boldsymbol{r}_{ij}}{R_{ij} R_{ik}} \cdot \frac{\partial \boldsymbol{r}_{ik}}{\partial \boldsymbol{r}_m} - \frac{\boldsymbol{r}_{ij} \cdot \boldsymbol{r}_{ik}}{R_{ij}^2 R_{ik}}  \frac{\partial R_{ij}}{\partial \boldsymbol{r}_m} - \frac{\boldsymbol{r}_{ij} \cdot \boldsymbol{r}_{ik}}{R_{ij} R_{ik}^2} \frac{\partial R_{ik}}{\partial \boldsymbol{r}_m}
\end{equation}
\begin{equation}
    \frac{\partial \boldsymbol{r}_{ij}}{\partial \boldsymbol{r}_m} = 
    \begin{bmatrix} 
        \delta_{mj} - \delta_{mi} & 0 & 0 \\
        0 & \delta_{mj} - \delta_{mi} & 0 \\
        0 & 0 & \delta_{mj} - \delta_{mi} \\
    \end{bmatrix}
\end{equation}
where $\delta_{mj}$ is the Kronecker delta between atom $m$ and $j$.

\section{The Derivatives of EAD}\label{dEAD}
The expression of the derivative with respect to an interacting atom $m$ is shown in the following:
\begin{equation}\label{ead_derivative}
    \frac{\partial \rho_i}{\partial \boldsymbol{r}_m} = \sum_{l_x, l_y, l_z=0}^{l_x+l_y+l_z=L} \frac{2L_{\max}!}{l_x!l_y!l_z!} \bigg[\sum_{j\neq i}^{N} Z_j \Phi\bigg] \bigg[\sum_{j\neq i}^{N} Z_j \frac{\partial \Phi}{\partial \boldsymbol{r}_m}\bigg]
\end{equation}
where the derivative of $\Phi$ with respect to an interacting atom $m$ is
\begin{equation}
    \begin{split}
        \frac{\partial \Phi}{\partial \boldsymbol{r}_m} = \frac{e^{-\eta(R_{ij}-R_s)^2}}{R_c^{l_x+l_y+l_z}} \bigg[\bigg(\frac{\partial x^{l_x}_{ij}}{\partial \boldsymbol{r}_m}  y^{l_y}_{ij} z^{l_z}_{ij} + x^{l_x}_{ij} \frac{\partial y^{l_y}_{ij}}{\partial \boldsymbol{r}_m} z^{l_z}_{ij} + x^{l_x}_{ij} y^{l_y}_{ij} \frac{\partial z^{l_z}_{ij}}{\partial \boldsymbol{r}_m} \bigg) f_c \\
        + x^{l_x}_{ij} y^{l_y}_{ij} z^{l_z}_{ij} \bigg(\frac{\partial f_c}{\partial R_{ij}} - 2 f_c \eta(R_{ij}-R_s)\bigg)\frac{\partial R_{ij}}{\partial \boldsymbol{r}_m} \bigg)\bigg]
    \end{split}
\end{equation}

\bibliographystyle{elsarticle-num}
\bibliography{ref.bib}

\begin{thebibliography}{10}
\expandafter\ifx\csname url\endcsname\relax
  \def\url#1{\texttt{#1}}\fi
\expandafter\ifx\csname urlprefix\endcsname\relax\def\urlprefix{URL }\fi
\expandafter\ifx\csname href\endcsname\relax
  \def\href#1#2{#2} \def\path#1{#1}\fi

\bibitem{yamakov2002dislocation}
V.~Yamakov, D.~Wolf, S.~R. Phillpot, A.~K. Mukherjee, H.~Gleiter, Dislocation
  processes in the deformation of nanocrystalline aluminium by
  molecular-dynamics simulation, Nat. Mater. 1~(1) (2002) 45--49.
\newblock \href {http://dx.doi.org/https://doi.org/10.1038/nmat700}
  {\path{doi:https://doi.org/10.1038/nmat700}}.

\bibitem{terrones2002molecular}
M.~Terrones, F.~Banhart, N.~Grobert, J.-C. Charlier, H.~Terrones, P.~Ajayan,
  Molecular junctions by joining single-walled carbon nanotubes, Phys. Rev.
  Lett. 89~(7) (2002) 075505.
\newblock \href
  {http://dx.doi.org/https://doi.org/10.1103/PhysRevLett.89.075505}
  {\path{doi:https://doi.org/10.1103/PhysRevLett.89.075505}}.

\bibitem{li2010dislocation}
X.~Li, Y.~Wei, L.~Lu, K.~Lu, H.~Gao, Dislocation nucleation governed softening
  and maximum strength in nano-twinned metals, Nature 464~(7290) (2010)
  877--880.
\newblock \href {http://dx.doi.org/https://doi.org/10.1038/nature08929}
  {\path{doi:https://doi.org/10.1038/nature08929}}.

\bibitem{kresse1993ab}
G.~Kresse, J.~Hafner, Ab initio molecular dynamics for liquid metals, Phys.
  Rev. B 47~(1) (1993) 558.
\newblock \href {http://dx.doi.org/https://doi.org/10.1103/PhysRevB.47.558}
  {\path{doi:https://doi.org/10.1103/PhysRevB.47.558}}.

\bibitem{kohn1965self}
W.~Kohn, L.~J. Sham, Self-consistent equations including exchange and
  correlation effects, Phys. Rev. 140~(4A) (1965) A1133.
\newblock \href {http://dx.doi.org/https://doi.org/10.1103/PhysRev.140.A1133}
  {\path{doi:https://doi.org/10.1103/PhysRev.140.A1133}}.

\bibitem{daw1984embedded}
M.~S. Daw, M.~I. Baskes, Embedded-atom method: Derivation and application to
  impurities, surfaces, and other defects in metals, Phys. Rev. B 29~(12)
  (1984) 6443.
\newblock \href {http://dx.doi.org/https://doi.org/10.1103/PhysRevB.29.6443}
  {\path{doi:https://doi.org/10.1103/PhysRevB.29.6443}}.

\bibitem{daw1993embedded}
M.~S. Daw, S.~M. Foiles, M.~I. Baskes, The embedded-atom method: a review of
  theory and applications, Materials Science Reports 9~(7-8) (1993) 251--310.
\newblock \href
  {http://dx.doi.org/https://doi.org/10.1016/0920-2307(93)90001-U}
  {\path{doi:https://doi.org/10.1016/0920-2307(93)90001-U}}.

\bibitem{tersoff1986new}
J.~Tersoff, New empirical model for the structural properties of silicon, Phys.
  Rev. Lett. 56~(6) (1986) 632.
\newblock \href {http://dx.doi.org/https://doi.org/10.1103/PhysRevLett.56.632}
  {\path{doi:https://doi.org/10.1103/PhysRevLett.56.632}}.

\bibitem{stillinger1985computer}
F.~H. Stillinger, T.~A. Weber, Computer simulation of local order in condensed
  phases of silicon, Phys. Rev. B 31~(8) (1985) 5262.
\newblock \href {http://dx.doi.org/https://doi.org/10.1103/PhysRevB.31.5262}
  {\path{doi:https://doi.org/10.1103/PhysRevB.31.5262}}.

\bibitem{mackerell1998all}
A.~D. MacKerell~Jr, D.~Bashford, M.~Bellott, R.~L. Dunbrack~Jr, J.~D. Evanseck,
  M.~J. Field, S.~Fischer, J.~Gao, H.~Guo, S.~Ha, et~al., All-atom empirical
  potential for molecular modeling and dynamics studies of proteins, J. Phys.
  Chem. B 102~(18) (1998) 3586--3616.
\newblock \href {http://dx.doi.org/https://doi.org/10.1021/jp973084f}
  {\path{doi:https://doi.org/10.1021/jp973084f}}.

\bibitem{behler2015constructing}
J.~Behler, Constructing high-dimensional neural network potentials: A tutorial
  review, International Journal of Quantum Chemistry 115~(16) (2015)
  1032--1050.
\newblock \href {http://dx.doi.org/https://doi.org/10.1002/qua.24890}
  {\path{doi:https://doi.org/10.1002/qua.24890}}.

\bibitem{artrith2012high}
N.~Artrith, J.~Behler, High-dimensional neural network potentials for metal
  surfaces: A prototype study for copper, Phys. Rev. B 85~(4) (2012) 045439.

\bibitem{li2017study}
W.~Li, Y.~Ando, E.~Minamitani, S.~Watanabe, Study of {Li} atom diffusion in
  amorphous {$\textrm{Li}_\textrm{3}\textrm{PO}_\textrm{4}$} with neural
  network potential, J. Chem. Phys. 147~(21) (2017) 214106.
\newblock \href {http://dx.doi.org/https://doi.org/10.1063/1.4997242}
  {\path{doi:https://doi.org/10.1063/1.4997242}}.

\bibitem{behler2007generalized}
J.~Behler, M.~Parrinello, Generalized neural-network representation of
  high-dimensional potential-energy surfaces, Phys. Rev. Lett. 98~(14) (2007)
  146401.
\newblock \href
  {http://dx.doi.org/https://doi.org/10.1103/PhysRevLett.98.146401}
  {\path{doi:https://doi.org/10.1103/PhysRevLett.98.146401}}.

\bibitem{behler2011atom}
J.~Behler, Atom-centered symmetry functions for constructing high-dimensional
  neural network potentials, J. Chem. Phys. 134~(7) (2011) 074106.
\newblock \href {http://dx.doi.org/https://doi.org/10.1063/1.3553717}
  {\path{doi:https://doi.org/10.1063/1.3553717}}.

\bibitem{artrith2011high}
N.~Artrith, T.~Morawietz, J.~Behler, High-dimensional neural-network potentials
  for multicomponent systems: Applications to zinc oxide, Phys. Rev. B 83~(15)
  (2011) 153101.
\newblock \href {http://dx.doi.org/https://doi.org/10.1103/PhysRevB.83.153101}
  {\path{doi:https://doi.org/10.1103/PhysRevB.83.153101}}.

\bibitem{hajinazar2017stratified}
S.~Hajinazar, J.~Shao, A.~N. Kolmogorov, Stratified construction of neural
  network based interatomic models for multicomponent materials, Phys. Rev. B
  95~(1) (2017) 014114.
\newblock \href {http://dx.doi.org/https://doi.org/10.1103/PhysRevB.95.014114}
  {\path{doi:https://doi.org/10.1103/PhysRevB.95.014114}}.

\bibitem{gastegger2015high}
M.~Gastegger, P.~Marquetand, High-dimensional neural network potentials for
  organic reactions and an improved training algorithm, Journal of Chemical
  Theory and Computation 11~(5) (2015) 2187--2198.
\newblock \href {http://dx.doi.org/https://doi.org/10.1021/acs.jctc.5b00211}
  {\path{doi:https://doi.org/10.1021/acs.jctc.5b00211}}.

\bibitem{bartok2010gaussian}
A.~P. Bart{\'o}k, M.~C. Payne, R.~Kondor, G.~Cs{\'a}nyi, Gaussian approximation
  potentials: The accuracy of quantum mechanics, without the electrons, Phys.
  Rev. Lett. 104~(13) (2010) 136403.
\newblock \href
  {http://dx.doi.org/https://doi.org/10.1103/PhysRevLett.104.136403}
  {\path{doi:https://doi.org/10.1103/PhysRevLett.104.136403}}.

\bibitem{bartok2013representing}
A.~P. Bart{\'o}k, R.~Kondor, G.~Cs{\'a}nyi, On representing chemical
  environments, Phys. Rev. B 87~(18) (2013) 184115.
\newblock \href {http://dx.doi.org/https://doi.org/10.1103/PhysRevB.87.184115}
  {\path{doi:https://doi.org/10.1103/PhysRevB.87.184115}}.

\bibitem{thompson2015spectral}
A.~P. Thompson, L.~P. Swiler, C.~R. Trott, S.~M. Foiles, G.~J. Tucker, Spectral
  neighbor analysis method for automated generation of quantum-accurate
  interatomic potentials, J. Comput. Phys. 285 (2015) 316--330.
\newblock \href {http://dx.doi.org/https://doi.org/10.1016/j.jcp.2014.12.018}
  {\path{doi:https://doi.org/10.1016/j.jcp.2014.12.018}}.

\bibitem{shapeev2016moment}
A.~V. Shapeev, Moment tensor potentials: A class of systematically improvable
  interatomic potentials, Multiscale Modeling \& Simulation 14~(3) (2016)
  1153--1173.
\newblock \href {http://dx.doi.org/https://doi.org/10.1137/15M1054183}
  {\path{doi:https://doi.org/10.1137/15M1054183}}.

\bibitem{chen2017accurate}
C.~Chen, Z.~Deng, R.~Tran, H.~Tang, I.-H. Chu, S.~P. Ong, Accurate force field
  for molybdenum by machine learning large materials data, Phys. Rev. Mater.
  1~(4) (2017) 043603.
\newblock \href
  {http://dx.doi.org/https://doi.org/10.1103/PhysRevMaterials.1.043603}
  {\path{doi:https://doi.org/10.1103/PhysRevMaterials.1.043603}}.

\bibitem{li2018quantum}
X.-G. Li, C.~Hu, C.~Chen, Z.~Deng, J.~Luo, S.~P. Ong, Quantum-accurate spectral
  neighbor analysis potential models for ni-mo binary alloys and fcc metals,
  Phys. Rev. B 98~(9) (2018) 094104.
\newblock \href {http://dx.doi.org/https://doi.org/10.1103/PhysRevB.98.094104}
  {\path{doi:https://doi.org/10.1103/PhysRevB.98.094104}}.

\bibitem{szlachta2014accuracy}
W.~J. Szlachta, A.~P. Bart{\'o}k, G.~Cs{\'a}nyi, Accuracy and transferability
  of gaussian approximation potential models for tungsten, Phys. Rev. B 90~(10)
  (2014) 104108.
\newblock \href {http://dx.doi.org/https://doi.org/10.1103/PhysRevB.90.104108}
  {\path{doi:https://doi.org/10.1103/PhysRevB.90.104108}}.

\bibitem{deringer2018data1}
V.~L. Deringer, D.~M. Proserpio, G.~Cs{\'a}nyi, C.~J. Pickard, Data-driven
  learning and prediction of inorganic crystal structures, Faraday discussions
  211 (2018) 45--59.
\newblock \href {http://dx.doi.org/https://doi.org/10.1039/C8FD00034D}
  {\path{doi:https://doi.org/10.1039/C8FD00034D}}.

\bibitem{deringer2018data2}
V.~L. Deringer, C.~J. Pickard, G.~Cs{\'a}nyi, Data-driven learning of total and
  local energies in elemental boron, Phys. Rev. Lett. 120~(15) (2018) 156001.
\newblock \href
  {http://dx.doi.org/https://doi.org/10.1103/PhysRevLett.120.156001}
  {\path{doi:https://doi.org/10.1103/PhysRevLett.120.156001}}.

\bibitem{podryabinkin2019accelerating}
E.~V. Podryabinkin, E.~V. Tikhonov, A.~V. Shapeev, A.~R. Oganov, Accelerating
  crystal structure prediction by machine-learning interatomic potentials with
  active learning, Phys. Rev. B 99~(6) (2019) 064114.
\newblock \href {http://dx.doi.org/10.1103/PhysRevB.99.064114}
  {\path{doi:10.1103/PhysRevB.99.064114}}.

\bibitem{singraber2019parallel}
A.~Singraber, T.~Morawietz, J.~Behler, C.~Dellago, Parallel multistream
  training of high-dimensional neural network potentials, Journal of chemical
  theory and computation 15~(5) (2019) 3075--3092.
\newblock \href {http://dx.doi.org/https://doi.org/10.1021/acs.jctc.8b01092}
  {\path{doi:https://doi.org/10.1021/acs.jctc.8b01092}}.

\bibitem{lee2019simple}
K.~Lee, D.~Yoo, W.~Jeong, S.~Han, Simple-nn: An efficient package for training
  and executing neural-network interatomic potentials, Computer Physics
  Communications 242 (2019) 95--103.
\newblock \href {http://dx.doi.org/https://doi.org/10.1016/j.cpc.2019.04.014}
  {\path{doi:https://doi.org/10.1016/j.cpc.2019.04.014}}.

\bibitem{khorshidi2016amp}
A.~Khorshidi, A.~A. Peterson, Amp: A modular approach to machine learning in
  atomistic simulations, Computer Physics Communications 207 (2016) 310--324.
\newblock \href {http://dx.doi.org/https://doi.org/10.1016/j.cpc.2016.05.010}
  {\path{doi:https://doi.org/10.1016/j.cpc.2016.05.010}}.

\bibitem{PINN-2020}
Y.~Shao, M.~Hellström, P.~D. Mitev, L.~Knijff, C.~Zhang,
  \href{https://doi.org/10.1021/acs.jcim.9b00994}{Pinn: A python library for
  building atomic neural networks of molecules and materials}, Journal of
  Chemical Information and Modeling 60~(3) (2020) 1184--1193, pMID: 31935100.
\newblock \href {http://arxiv.org/abs/https://doi.org/10.1021/acs.jcim.9b00994}
  {\path{arXiv:https://doi.org/10.1021/acs.jcim.9b00994}}, \href
  {http://dx.doi.org/10.1021/acs.jcim.9b00994}
  {\path{doi:10.1021/acs.jcim.9b00994}}.
\newline\urlprefix\url{https://doi.org/10.1021/acs.jcim.9b00994}

\bibitem{schutt2018schnet}
K.~T. Sch{\"u}tt, H.~E. Sauceda, P.-J. Kindermans, A.~Tkatchenko, K.-R.
  M{\"u}ller, Schnet--a deep learning architecture for molecules and materials,
  J. Chem. Phys. 148~(24) (2018) 241722.
\newblock \href {http://dx.doi.org/https://doi.org/10.1063/1.5019779}
  {\path{doi:https://doi.org/10.1063/1.5019779}}.

\bibitem{artrith2016implementation}
N.~Artrith, A.~Urban, An implementation of artificial neural-network potentials
  for atomistic materials simulations: Performance for tio2, Computational
  Materials Science 114 (2016) 135--150.
\newblock \href
  {http://dx.doi.org/https://doi.org/10.1016/j.commatsci.2015.11.047}
  {\path{doi:https://doi.org/10.1016/j.commatsci.2015.11.047}}.

\bibitem{yanxon2020transferability}
H.~Yanxon, D.~Zagaceta, B.~C. Wood, Q.~Zhu, Neural networks potential from the
  bispectrum component: A case study on crystalline silicon, arXiv preprint
  arXiv:2001.00972.

\bibitem{Zagaceta2020SpectralNN}
D.~Zagaceta, H.~Yanxon, Q.~Zhu, Spectral neural network potentials for binary
  alloys, arXiv: Computational Physics.

\bibitem{zhang2018end}
L.~Zhang, J.~Han, H.~Wang, W.~Saidi, R.~Car, E.~Weinan, End-to-end symmetry
  preserving inter-atomic potential energy model for finite and extended
  systems, in: Advances in Neural Information Processing Systems, 2018, pp.
  4436--4446.

\bibitem{zhang2019embedded}
Y.~Zhang, C.~Hu, B.~Jiang, Embedded atom neural network potentials: Efficient
  and accurate machine learning with a physically inspired representation, J.
  Phys. Chem. Lett. 10~(17) (2019) 4962--4967.
\newblock \href {http://dx.doi.org/https://doi.org/10.1021/acs.jpclett.9b02037}
  {\path{doi:https://doi.org/10.1021/acs.jpclett.9b02037}}.

\bibitem{li2019unravelling}
X.-G. Li, C.~Chen, H.~Zheng, S.~P. Ong, Unravelling complex strengthening
  mechanisms in the {NbMoTaW} multi-principal element alloy with machine
  learning potentials, arXiv preprint arXiv:1912.01789\href
  {http://dx.doi.org/https://arxiv.org/abs/1912.01789}
  {\path{doi:https://arxiv.org/abs/1912.01789}}.

\bibitem{HIMANEN2020106949}
Dscribe: Library of descriptors for machine learning in materials science,
  Computer Physics Communications 247 (2020) 106949.
\newblock \href {http://dx.doi.org/https://doi.org/10.1016/j.cpc.2019.106949}
  {\path{doi:https://doi.org/10.1016/j.cpc.2019.106949}}.

\bibitem{rupp2012fast}
M.~Rupp, A.~Tkatchenko, K.-R. M{\"u}ller, O.~A. Von~Lilienfeld, Fast and
  accurate modeling of molecular atomization energies with machine learning,
  Physical review letters 108~(5) (2012) 058301.
\newblock \href
  {http://dx.doi.org/https://doi.org/10.1103/PhysRevLett.108.058301}
  {\path{doi:https://doi.org/10.1103/PhysRevLett.108.058301}}.

\bibitem{faber2015crystal}
F.~Faber, A.~Lindmaa, O.~A. von Lilienfeld, R.~Armiento, Crystal structure
  representations for machine learning models of formation energies,
  International Journal of Quantum Chemistry 115~(16) (2015) 1094--1101.
\newblock \href {http://dx.doi.org/https://doi.org/10.1002/qua.24917}
  {\path{doi:https://doi.org/10.1002/qua.24917}}.

\bibitem{huo2017unified}
H.~Huo, M.~Rupp, Unified representation of molecules and crystals for machine
  learning, arXiv preprint arXiv:1704.06439.

\bibitem{gastegger2018wacsf}
M.~Gastegger, L.~Schwiedrzik, M.~Bittermann, F.~Berzsenyi, P.~Marquetand,
  wacsf—weighted atom-centered symmetry functions as descriptors in machine
  learning potentials, J. Chem. Phys. 148~(24) (2018) 241709.
\newblock \href {http://dx.doi.org/https://doi.org/10.1063/1.5019667}
  {\path{doi:https://doi.org/10.1063/1.5019667}}.

\bibitem{daw1983semiempirical}
M.~S. Daw, M.~I. Baskes, Semiempirical, quantum mechanical calculation of
  hydrogen embrittlement in metals, Phys. Rev. Lett. 50~(17) (1983) 1285.
\newblock \href {http://dx.doi.org/https://doi.org/10.1103/PhysRevLett.50.1285}
  {\path{doi:https://doi.org/10.1103/PhysRevLett.50.1285}}.

\bibitem{larsen2017atomic}
A.~H. Larsen, J.~J. Mortensen, J.~Blomqvist, I.~E. Castelli, R.~Christensen,
  M.~Du{\l}ak, J.~Friis, M.~N. Groves, B.~Hammer, C.~Hargus, et~al., The atomic
  simulation environment—a python library for working with atoms, Journal of
  Physics: Condensed Matter 29~(27) (2017) 273002.
\newblock \href {http://dx.doi.org/https://doi.org/10.1088/1361-648X/aa680e}
  {\path{doi:https://doi.org/10.1088/1361-648X/aa680e}}.

\bibitem{walt2011numpy}
S.~v.~d. Walt, S.~C. Colbert, G.~Varoquaux, The numpy array: a structure for
  efficient numerical computation, Computing in Science \& Engineering 13~(2)
  (2011) 22--30.
\newblock \href {http://dx.doi.org/https://doi.org/10.1109/MCSE.2011.37}
  {\path{doi:https://doi.org/10.1109/MCSE.2011.37}}.

\bibitem{NEURIPS2019_9015_torch}
A.~Paszke, S.~Gross, F.~Massa, A.~Lerer, J.~Bradbury, G.~Chanan, T.~Killeen,
  Z.~Lin, N.~Gimelshein, L.~Antiga, A.~Desmaison, A.~Kopf, E.~Yang, Z.~DeVito,
  M.~Raison, A.~Tejani, S.~Chilamkurthy, B.~Steiner, L.~Fang, J.~Bai,
  S.~Chintala, Pytorch: An imperative style, high-performance deep learning
  library, in: H.~Wallach, H.~Larochelle, A.~Beygelzimer, F.~dAlche-Buc, E.~Fox, R.~Garnett (Eds.), Advances in Neural Information
  Processing Systems 32, Curran Associates, Inc., 2019, pp. 8024--8035.

\bibitem{paszke2017automatic}
A.~Paszke, S.~Gross, S.~Chintala, G.~Chanan, E.~Yang, Z.~DeVito, Z.~Lin,
  A.~Desmaison, L.~Antiga, A.~Lerer, Automatic differentiation in pytorch.

\bibitem{liu1989limited}
D.~C. Liu, J.~Nocedal, On the limited memory {BFGS} method for large scale
  optimization, Mathematical programming 45~(1-3) (1989) 503--528.
\newblock \href {http://dx.doi.org/https://doi.org/10.1007/BF01589116}
  {\path{doi:https://doi.org/10.1007/BF01589116}}.

\bibitem{kingma2014adam}
D.~P. Kingma, J.~Ba, Adam: A method for stochastic optimization, arXiv preprint
  arXiv:1412.6980\href {http://dx.doi.org/https://arxiv.org/abs/1412.6980}
  {\path{doi:https://arxiv.org/abs/1412.6980}}.

\bibitem{qian1999momentum}
N.~Qian, On the momentum term in gradient descent learning algorithms, Neural
  networks 12~(1) (1999) 145--151.
\newblock \href
  {http://dx.doi.org/https://doi.org/10.1016/S0893-6080(98)00116-6}
  {\path{doi:https://doi.org/10.1016/S0893-6080(98)00116-6}}.

\bibitem{LAMMPS}
S.~Plimpton, Fast parallel algorithms for short-range molecular dynamics, J.
  Comp. Phys. 117~(1) (1995) 1 -- 19.
\newblock \href {http://dx.doi.org/https://doi.org/10.1006/jcph.1995.1039}
  {\path{doi:https://doi.org/10.1006/jcph.1995.1039}}.

\bibitem{phonopy}
A.~Togo, I.~Tanaka, First principles phonon calculations in materials science,
  Scr. Mater. 108 (2015) 1--5.

\bibitem{HINUMA2017140}
Y.~Hinuma, G.~Pizzi, Y.~Kumagai, F.~Oba, I.~Tanaka, Band structure diagram
  paths based on crystallography, Comput. Mater. Sci. 128 (2017) 140 -- 184.
\newblock \href
  {http://dx.doi.org/https://doi.org/10.1016/j.commatsci.2016.10.015}
  {\path{doi:https://doi.org/10.1016/j.commatsci.2016.10.015}}.

\bibitem{matscipy}
Matscipy, \url{https://gitlab.com/libAtoms/matscipy}.

\bibitem{Vasp-PRB-1996}
G.~Kresse, J.~Furthm\"uller, Efficient iterative schemes for ab initio
  total-energy calculations using a plane-wave basis set, Phys. Rev. B 54
  (1996) 11169--11186.
\newblock \href {http://dx.doi.org/10.1103/PhysRevB.54.11169}
  {\path{doi:10.1103/PhysRevB.54.11169}}.

\bibitem{PBE-PRL-1996}
J.~P. Perdew, K.~Burke, M.~Ernzerhof, Generalized gradient approximation made
  simple, Phys. Rev. Lett. 77 (1996) 3865--3868.
\newblock \href {http://dx.doi.org/10.1103/PhysRevLett.77.3865}
  {\path{doi:10.1103/PhysRevLett.77.3865}}.

\bibitem{yeh2004nanostructured}
J.-W. Yeh, S.-K. Chen, S.-J. Lin, J.-Y. Gan, T.-S. Chin, T.-T. Shun, C.-H.
  Tsau, S.-Y. Chang, Nanostructured high-entropy alloys with multiple principal
  elements: novel alloy design concepts and outcomes, Advanced Engineering
  Materials 6~(5) (2004) 299--303.
\newblock \href {http://dx.doi.org/https://doi.org/10.1002/adem.200300567}
  {\path{doi:https://doi.org/10.1002/adem.200300567}}.

\bibitem{senkov2011mechanical}
O.~N. Senkov, G.~Wilks, J.~Scott, D.~B. Miracle, Mechanical properties of
  $\textrm{Nb}_\textrm{25}\textrm{Mo}_\textrm{25}\textrm{Ta}_\textrm{25}\textrm{W}_\textrm{25}$
  and
  $\textrm{V}_\textrm{20}\textrm{Nb}_\textrm{20}\textrm{Mo}_\textrm{20}\textrm{Ta}_\textrm{20}\textrm{W}_\textrm{20}$
  refractory high entropy alloys, Intermetallics 19~(5) (2011) 698--706.
\newblock \href
  {http://dx.doi.org/https://doi.org/10.1016/j.intermet.2011.01.004}
  {\path{doi:https://doi.org/10.1016/j.intermet.2011.01.004}}.

\bibitem{MP-2013}
A.~Jain, S.~P. Ong, G.~Hautier, W.~Chen, W.~D. Richards, S.~Dacek, S.~Cholia,
  D.~Gunter, D.~Skinner, G.~Ceder, et~al., Commentary: The materials project: A
  materials genome approach to accelerating materials innovation, Apl Materials
  1~(1) (2013) 011002.
\newblock \href {http://dx.doi.org/10.1063/1.4812323}
  {\path{doi:10.1063/1.4812323}}.

\end{thebibliography}


\begin{thebibliography}{0}
\bibitem{1} \url{https://opensource.org/licenses/MIT}

\end{thebibliography}
\end{document}